\documentclass[aps,pra,notitlepage,twocolumn,10pt,a4paper]{revtex4-1}

\usepackage{xcolor,graphicx,ulem,soul}
\usepackage{amsmath,amssymb,esint}
\usepackage[colorlinks=true,urlcolor=blue,citecolor=red,linkcolor=blue]{hyperref}
\usepackage{braket}
\usepackage{cancel}
\usepackage{comment}
\usepackage[utf8]{inputenc}

\newcommand{\w}{\omega}
\newcommand{\W}{\Omega}
\newcommand{\cdag}{\hat{c}^{\dagger}}
\newcommand{\dddag}{\hat{d}^{\dagger}}
\newcommand{\xdag}{\hat{x}^{\dagger}}
\newcommand{\ydag}{\hat{y}^{\dagger}}
\newcommand{\adag}{\hat{a}^{\dagger}}
\newcommand{\bdag}{\hat{b}^{\dagger}}
\newcommand{\vac}{\ket{\text{vac}}}
\newcommand{\ud}{\textrm{d}}
\newcommand{\modsqr}[1]{\left| #1 \right|^2}
\newcommand{\nn}{\nonumber}

\begin{document}


\title{Spectrally-resolved four-photon interference of time-frequency entangled photons}


\author{Sofiane Merkouche$^{*1}$, Val\'{e}rian Thiel$^1$, 
and Brian J. Smith$^{1}$}
\email[]{Corresponding author: oqt@uoregon.edu}
\affiliation{$^1$ Department of Physics and Oregon Center for Optical, Molecular, and Quantum Science, University of Oregon, Eugene, Oregon 97403, USA}


\date{\today}

\begin{abstract}
Pairs of photons entangled in their time-frequency degree of freedom are of great interest in quantum optics research and applications, due to their relative ease of generation and their high capacity for encoding information. Here we analyze, both theoretically and experimentally, the behavior of phase-insensitive spectrally-resolved interferences arising from two pairs of time-frequency entangled photons. At its core, this is a multimode entanglement swapping experiment, whereby a spectrally resolved joint measurement on the idler photons from both pairs results in projecting the signal photons onto a Bell state whose form depends on the measurement outcome. Our analysis is a thorough exploration of what can be achieved using time-frequency entanglement and spectrally-resolved Bell-state measurements.
\end{abstract}


\maketitle

\section{Introduction}

With the advent of the quantum information age, it is well-established by now that the encoding of quantum information into the degrees of freedom of light is the key component of quantum communication networks \cite{Kimble2008}. Whereas polarization and spatial-mode encoding benefit from ease of implementation, they are prone to scrambling from environmental noise and optical-fiber transmission, which undermines their suitability for long-distance communication. Meanwhile, the time-frequency (TF) degree of freedom is more robust in this regard (distinct colors stay distinct), and TF-encoding has long been the standard for classical communication for this reason \cite{Brecht2015}. 

In the quantum regime, entanglement plays a key role in many protocols for computation \cite{Jozsa2003} and communication \cite{Ekert1991, Riedmatten2005}. Furthermore, entanglement of photons has recently been of great interest to the metrology and spectroscopy communities due to its promise of enhancements in sensitivity beyond what is attainable in the classical domain \cite{Raymer2013}. It comes as no surprise, then, that the generation of pairs of photons in well-defined TF-entangled states is a widely-researched area of quantum optics, and great strides have been made over this terrain over the past two decades \cite{Ansari2018}. In addition to state generation, harnessing the full capabilities of quantum entanglement also requires the ability to projectively measure onto entangled states. Indeed, such entangled measurements, of which the Bell-state measurement is the prototype \cite{Vertesi2011}, are nearly as ubiquitous in quantum protocols as entangled states, most notably in quantum teleportation \cite{Bennett1993} and entanglement swapping \cite{Zukowski1993}.

In this work, we describe theoretically, and demonstrate experimentally, a novel and versatile TF entanglement-swapping scheme. Our scheme relies on the multimode nature of TF-entanglement in pairs of photons generated in SPDC, commonly denoted signal and idler. The central component of the setup is a multimode, frequency-resolved Bell-state measurement (BSM), performed on idler photons from two independent pairs. The BSM heralds the surviving signal photons onto a pulsed Bell state whose central frequencies depend on the result of the BSM. In this way we are able to herald multiple orthogonal Bell pairs and verify entanglement in each pair, all derived from the same source state and measurement scheme. In section \ref{sec:theory} we outline the theory underlying our work. In section \ref{sec:experiments} we describe the experimental setup. Then, in section \ref{sec:results} we describe our results, which show remarkable agreement with a simple and intuitive Gaussian model using pure quantum states, before concluding in section \ref{sec:conclusion}. Finally, in the Appendix we cover the more technical details of our work which would otherwise encumber the account of our main results.

\section{Theory} \label{sec:theory}

\subsection{Four photon state} \label{sec:4photons}

\begin{figure}[t]
	\centering
	\includegraphics[width=\linewidth]{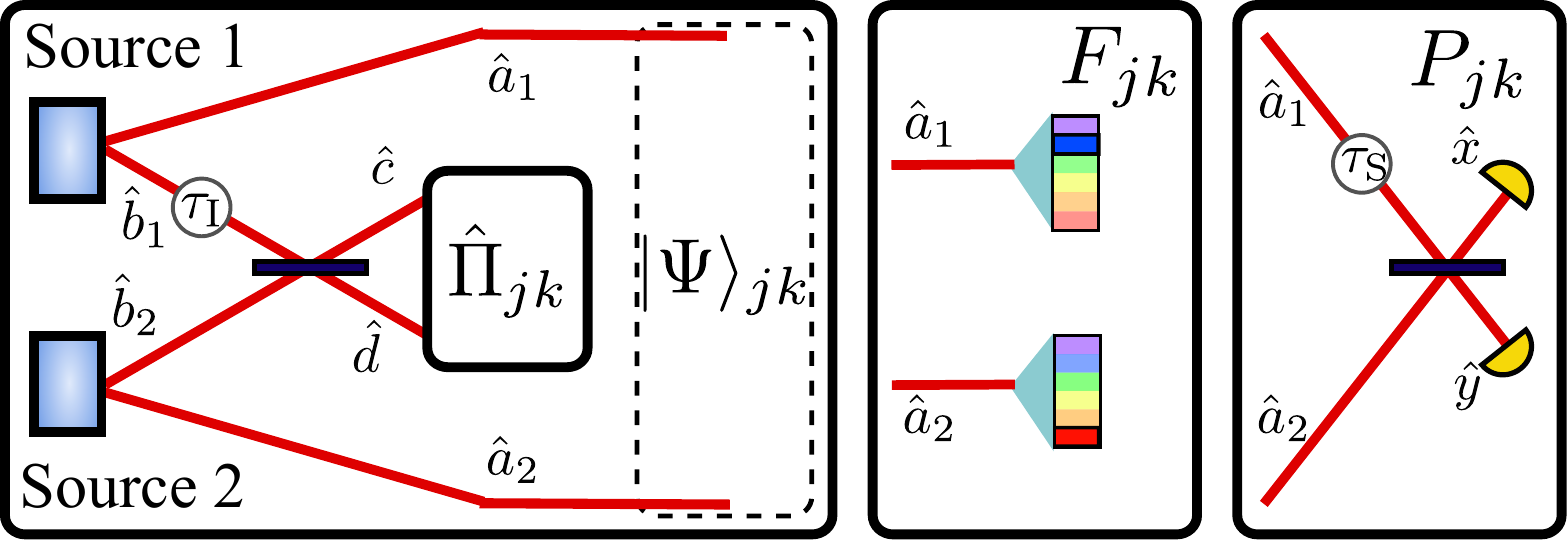}
	\caption{Conceptual scheme of the experiment. Two sources, 1 and 2, emit photon pairs, labelled $\hat{a}$ and $\hat{b}$  for the signal and idler photons, respectively. The photon pairs from the first source undergo a delay $\tau$ relative to their counterpart on source 1.}
	\label{fig:supp:simple:scheme}
\end{figure}

The entanglement swapping setup for this work is depicted schematically in Fig. \ref{fig:supp:simple:scheme}. The setup consists of two independent spontaneous parametric down conversion (SPDC) sources, each of which generates pairs of photons into paths labeled by the bosonic operators $\hat{a}_n$ for the signal and $\hat{b}_n$ for the idler, where $n \in \{1,2\}$ labels the two sources. Here we consider pulsed collinear type II SPDC sources, where the signal and idler modes are distinguished through polarization, and all the light is collected into single-mode optical fibers, so that only the time-frequency of freedom is relevant.

In the low gain regime, the output state of $n^\textrm{th}$ SPDC source is given by
\begin{align}
	\ket{\psi_n}=\sum_{k=0}^\infty \frac{\sqrt{\eta_n}^k}{k!}\left(\int \ud\w_\text{S} \ud\w_\text{I} f_n(\w_\text{S}, \w_\text{I}) \adag_n(\w_\text{S}) \bdag_n(\w_\text{I})\right)^k\ket{\mathrm{vac}},
\end{align}
The function $f_n(\w_\text{S},\w_\text{I})$ is the joint spectral amplitude (JSA), given by
\begin{align}
    f_n(\w_\text{S},\w_\text{I}) = u_n(\w_\text{S}+\w_\text{I})\textrm{sinc}\left[ \frac{\Delta k_n(\w_\text{S},\w_\text{I}) L}{2}\right],
\end{align}
where $u_n$ represents the spectral mode function of the pump, $\Delta k_n$ is the wave-vector mismatch between the pump, signal and idler modes, and $L$ is the length of the interaction medium. Finally, $\eta_n$ is the gain of the parametric process, which depends on the length $L$, the non-linear strength of the material and the number of photons in the pump beam.

The SPDC state due to the two independent sources can be written as a tensor product $\ket{\psi_1}\otimes \ket{\psi_2}$. We expand this and keep only terms of order $\eta$, which are responsible for the four-photon contribution, obtaining the following normalized state:
\begin{equation}
\begin{gathered}
	\ket{\psi_\mathrm{\eta}} =\frac{\left(\sqrt{\eta_1\eta_2}\ket{\psi_{12}}+\frac{\eta_1}{2}\ket{\psi_{11}} + \frac{\eta_2}{2}\ket{\psi_{22}}\right)}{\sqrt{\eta_1\eta_2+\eta_1^2/4+\eta_2^2/4}}, \label{eq:psip}
\end{gathered}
\end{equation}
where
\begin{align}
    \ket{\psi_{nm}} = &\int \ud^4 \w \  f_n(\w_\text{S},\w_\text{I}) f_m(\w_\text{S}',\w_\text{I}') \times \nn \\ &\adag_n(\w_\text{S})\adag_m(\w_\text{S}')\bdag_n(\w_\text{I})\bdag_m(\w_\text{I}') \vac.
\end{align}
is the four-photon state arising from either a photon pair from each source, or two pairs from one source and none from the other. 

In order to facilitate the discussion of our results, we will make a few simplifying assumptions, and we will relax these assumptions whenever appropriate in the main text and discuss them further in the Appendix. First we will assume that both sources are identical, such that $f_1(\w_\text{S},\w_\text{I})=f_2(\w_\text{S},\w_\text{I})=f(\w_\text{S},\w_\text{I})$ and $\eta_1=\eta_2=\eta$. This is because in the experiment, the two sources are derived from double-pumping the same crystal as in Ref. \cite{Pan1998}, and are matched to a great degree as discussed in the experimental section. Further, any source mismatch does not reduce the quality of the entanglement in the swapped state, just the visibility of quantum interference in the method we use to verify entanglement. 

Second, we will assume that the relative phases between the three terms in \eqref{eq:psip} are random, that is, these terms are not mutually coherent do not contribute to any quantum interference. In reality there is such a coherence, which is due to the phase of the pump, and we observe this in both two- and four-photon interference, as we show in Appendix \ref{app:ccfringes}. However, this phase drifts over the course of the interference measurements we will describe, which are several hours long, and thus the $\ket{\psi_{11}}$ and $\ket{\psi_{22}}$ contribute to a constant background in these measurements.

Third, in the main text we will focus our attention solely on the $\ket{\psi_{12}}$ term of \eqref{eq:psip}, which corresponds to each source producing a pair of entangled photons. In most entanglement swapping and quantum teleportation experiments relying on SPDC sources, the other two terms (the double-pair terms), though present, do not contribute when the four photons are detected in coincidence, so that heralded states are \textit{post-selected} \cite{Wagenknecht2010}. Similarly, the double-pair terms only contribute in our setup in the aforementioned interference measurement, where they constitute a constant background which we measure and subtract.

Finally, as most of the experiments that we performed rely on performing a spectrally-resolved Bell state measurement (BSM) on the idler photon, we will assume for simplicity that our spectral resolution is unlimited. Although it introduces some slight abuses of notation, this limit has the critical benefit of providing a simple and intuitive model with which to understand the physics in terms of pure quantum states (the pure-state approximation). A more complete model taking into account the finite spectral resolution of the BSM and the resulting states is then easily constructed from this pure-state approximation in Appendix \ref{app:mixedstate}. As we shall see, the pure-state approximation is sufficient to account for the majority of the physics and the results of our experiment.


%

\subsection{Heralded state and JSI}

A Bell-state measurement is performed on the idler photons by resolving their frequencies at the output of a 50:50 beamsplitter, with a (small) path difference at the input giving a relative time delay $\tau_\text{I}$. The beamsplitter transformation is then defined by the relations
\begin{equation}
\begin{gathered}
	\cdag(\W) = \frac{ e^{i \W \tau_\text{I}}\bdag_1(\W) + \bdag_2(\W)}{\sqrt{2}},\\
	\dddag(\W') = \frac{e^{i \W' \tau_\text{I}}\bdag_1(\W') - \bdag_2(\W') }{\sqrt{2}},
\end{gathered}
\end{equation}
and coincidences are detected between $\hat{c}$ and $\hat{d}$. Note that for clarity, we will label as $\W$ the optical frequencies of the idlers and as $\w$ those of the signals. Because we use the pure state approximation, we can consider that this spectral measurement is achieved with perfect resolution, such that the POVM element for this detection is given by
\begin{equation}
\begin{gathered}
	\hat{\Pi}_{jk}^\textrm{BSM} = \ket{\W_j,\W_k}\bra{\W_j,\W_k},\\ \textrm{with }\ket{\W_j,\W_k}=\cdag(\W_j)\dddag(\W_k)\vac,
	\label{eq:supp:piBSMpure}
\end{gathered}
\end{equation}
which is a projector onto the monochromatic idler frequencies $\W_j$ and $\W_k$.
We then proceed to compute the heralded signal state, $\ket{\Psi_{jk}}$, defined by:
\begin{align}
	\ket{\Psi_{jk}}\otimes\ket{\W_j,\W_k}=\frac{\hat{\Pi}_{jk}^\textrm{BSM}\ket{\psi_{12}}}{\sqrt{p_{jk}}},
	\label{eq:supp:heraldedstatedeff}
\end{align}
where the norm $p_{jk}$ is given by
\begin{align}
	p_{jk}=\bra{\psi_{12}}\hat{\Pi}_{jk}^\textrm{BSM}\ket{\psi_{12}}.
\end{align}

To compute the form of $\ket{\Psi_{jk}}$, we will introduce, for convenience, the density matrices for the reduced states of the signal and idler photons, given respectively by,
\begin{equation}
\begin{gathered}
    \rho_\mathrm{S}(\w,\w')=\int \ud\W f(\w,\W) f^*(\w',\W),\\
    \rho_\mathrm{I}(\W,\W')=\int \ud\w f(\w,\W) f^*(\w,\W').
    \label{eq:rhoz}
\end{gathered}
\end{equation}
Since the sources are identical, the reduced density matrices obey the relation $\rho(\w,\w') = \rho^\ast(\w',\w)$. Using these, it is straightforward to show that
\begin{align}
    p_{jk} = \frac{1}{2}\Big[\rho_\mathrm{I}(\W_j,\W_j)\rho_\mathrm{I}(\W_k,\W_k)-|\rho_\mathrm{I}(\W_j,\W_k)|^2\cos{\theta_{jk}}\Big],
    \label{eq:pjk:theory}
\end{align}
with
\begin{align}
    \theta_{jk} = (\W_j-\W_k)\tau_\text{I}. \label{eq:thetajk}
\end{align}
This is the probability distribution for a coincidence between the idler photons at $(\W_j,\W_k)$, or equivalently, the JSI of the idlers at the output of the beamsplitter. 

Most importantly, the heralded signal Bell-state has the simple form
\begin{align}
	\ket{\Psi_{jk}}=
	\frac{\ket{\phi_j}_1\ket{\phi_k}_2 - e^{i\theta_{jk}}\ket{\phi_k}_1\ket{\phi_j}_2}{\sqrt{2\mathcal{C}_{jk}}},
	\label{eq:supp:heraldedstate:delay}
\end{align}
where $\ket{\phi_{j(k)}}_{1(2)}$ is a pulse-mode normalized single photon state \cite{Blow1990} given by
\begin{equation}
	\ket{\phi_{j(k)}}_{1(2)} = \int \ud \omega \, \phi_{j(k)}(\omega) \adag_{1(2)}(\omega)\vac, \label{eq:phiket}
\end{equation}
with
\begin{equation}
    \phi_{j(k)}(\omega) = \frac{f(\omega,\W_{j(k)})}{ \sqrt{\rho_\mathrm{I}(\W_{j(k)},\W_{j(k)})}}. \label{eq:phijk:theory}
\end{equation}
Essentially, $\ket{\phi_{j(k)}}$ is the state that the signal photon is projected onto, when its corresponding idler photon is detected at frequency $\W_{j(k)}$. Finally the normalization constant $\mathcal{C}_{jk}$ is given by
\begin{align}
    \mathcal{C}_{jk} = 1 - \modsqr{\braket{\phi_j|\phi_k}}\cos{\theta_{jk}}. \label{eq:psi:norm}
\end{align}
The functions $\phi_{j(k)}$ are defined from the JSA and several identities are shown in Appendix \ref{app:math}. Note that, while normalized, they are not necessarily orthogonal, hence the dependency on the modal overlap in the normalization from Eq.\eqref{eq:psi:norm}. In Sec.\ref{sec:results}, we will use a Gaussian approximation that gives a simple expression for those functions.

Since nearly all of our measurements are conditioned upon the BSM on the idler photons, we will use the state $\ket{\Psi_{jk}}$ to calculate our quantities of interest. In addition, we also consider the case of a non-resolving BSM, given by
\begin{align}
	\hat{\Pi}^\text{BSM} &=\sum_{j,k} \hat{\Pi}^\text{BSM}_{jk}\\
	\nonumber & =\int \ud\W_j\ud\W_k\ket{\W_j,\W_k}\bra{\W_j,\W_k},
\end{align}
which, upon taking $\text{Tr}\Big(\hat{\Pi}^\text{BSM}\ket{\psi_{12}}\bra{\psi_{12}}\Big)$, heralds the mixed state
\begin{equation}
	\hat{\rho}= \sum_{jk} p_{jk} \ket{\Psi_{jk}}\bra{\Psi_{jk}}, \label{eq:rhosum}
\end{equation}
and the quantities that arise from this state. Note that this state is not normalized, since $\sum_{jk}p_{jk} \neq 1$, but this fact is inconsequential to our measurements, as only the relative probabilities $p_{jk}$ are physically relevant. At this point we note that we are using the sum $\sum_{jk}$ loosely in place of the integral $\int \ud\W_j \ud\W_k$ because of the finite spectral resolution of our heralding BSM. This means that the pure state description of $\ket{\Psi_{jk}}$ is an approximation, but one whose validity we justify throughout the paper as well as in Appendix \ref{app:mixedstate} and also gives similar results as the model which takes into account the finite resolution of the BSM and the resultant impurity of the heralded state. 

\subsection{State characterization and entanglement verification}

We characterize the heralded state $\ket{\Psi_{jk}}$ first by measuring its joint spectral intensity (JSI). This measurement is defined by the POVM
\begin{equation}
\begin{gathered}
    \hat{\Pi}^\text{JSI}=\ket{\w_1,\w_2}\bra{\w_1,\w_2},\\
    \textrm{with }\ket{\w_1,\w_2} = \adag_1(\w_1)\adag_2(\w_2)\vac,
\end{gathered}
\end{equation}
and we calculate the resultant JSI as
\begin{align}
	F_{jk}&(\w_1,\w_2) = \bra{\Psi_{jk}}\hat{\Pi}^\text{JSI}\ket{\Psi_{jk}} = \left|  \Braket{\w_1, \w_2 | \Psi_{jk}} \right|^2. \label{eq:supp:heraldedJSIpure}
\end{align}
Using \eqref{eq:supp:heraldedstate:delay}, we obtain explicitly
\begin{align}
	F_{jk}(\w_1,\w_2) = \frac{1}{2 \mathcal{C}_{jk}} \Big| \phi_j(\w_1)\phi_k(\w_2) - e^{i\theta_{jk}}\phi_j(\w_2)\phi_k(\w_1) \Big|^2. \label{eq:supp:heraldedJSI:delay}
\end{align}
In the absence of spectral resolution in the BSM, the heralded state is $\hat{\rho}$ from Eq.\eqref{eq:rhosum}, and the measured JSI is given by
\begin{align}
	F(\w_1,\w_2) &= \text{Tr}\left(\hat{\rho}\ \hat{\Pi}^\text{JSI}\right) \nn \\
	&= \sum_{jk} p_{jk}\left|  \Braket{\w_1, \w_2 | \Psi_{jk}} \right|^2 \nn \\
	&= \sum_{jk} p_{jk} F_{jk}(\w_1,\w_2).
	\label{eq:supp:heraldedJSIsum}
\end{align}
In terms of previously defined quantities, this function takes the general form
\begin{align}
    F(\w_1,\w_2) = &\frac{1}{2}\Bigg[  \rho_\mathrm{S}(\w_1,\w_1)\rho_\mathrm{S}(\w_2,\w_2) \nn \\
    &- \left|\int \ud\W f(\w_1,\W)f^*(\w_2,\W)e^{i\W\tau_\text{I}}\right|^2 \Bigg]. \label{eq:heraldedJSI:full:delay}
\end{align}
To verify that the heralded state $\ket{\Psi_{jk}}$ is indeed entangled, beyond classical correlation, two-photon interference is used in a manner analogous to Refs. \cite{Ramelow2009,Graffitti2020}. Here the signal photons are combined at another 50:50 beamsplitter while scanning a relative delay between the input modes ($\hat{a}_1$ and $\hat{a}_2$) denoted by $\tau_\text{S}$ and monitoring coincidences at the output modes ($\hat{x}$ and $\hat{y}$) as depicted in Fig.~\ref{fig:supp:simple:scheme}. These modes transform as:
\begin{align}
	&\xdag(\omega) = \frac{e^{i\w \tau_\text{S}}\adag_1(\omega) + \adag_2(\omega)}{\sqrt{2}}, \nn \\
	&\ydag(\omega') = \frac{e^{i\w' \tau_\text{S}}\adag_1(\omega') - \adag_2(\omega')}{\sqrt{2}}.
\end{align}
The POVM associated with such a coincidence detection is defined as
\begin{align}
\begin{gathered}
	\hat{\Pi}_\textrm{verif} = \int \ud\w \ud \w'  \ket{\w,\w'}\bra{\w,\w'},\\
	\textrm{with }\ket{\w,\w'}=\xdag(\w)\ydag(\w')\vac,
	\label{eq:povm:verif}
\end{gathered}
\end{align}
and the probability of detecting a coincidence heralded for the input state $\ket{\Psi_{jk}}$ is given by
\begin{align}
	P_{jk}(\tau_\text{S}) &= \Braket{\Psi_{jk}| \hat{\Pi}_\textrm{verif} | \Psi_{jk}}
	\nn \\
	&= \int \ud\w\ud\w' \, \Big |\braket{\w , \w' | \Psi_{jk}} \Big|^2. \label{eq:supp:PjkPure}
\end{align}
Evaluating this using Eq. \eqref{eq:supp:heraldedstate:delay}, we obtain
\begin{align}
	&P_{jk}(\tau_\text{S}) = \frac{1}{2\mathcal{C}_{jk}} \Big(1 \nn \\ 
	&+\mathcal{F}_j(\tau_\text{S})\mathcal{F}_k(\tau_\text{S}) \cos{\left[(\w_j-\w_k)\tau_\text{S} - \theta_{jk}\right]} \nn \\ &-\mathcal{O}(\left|\braket{\phi_j|\phi_k}\right|^2)\Big). \label{eq:supp:PjkPure2}
\end{align}
%
%
where $\mathcal{F}_{j(k)}(\tau_\text{S})$ is the Fourier transform of $|\phi_{j(k)}(\omega)|^2$, $\omega_{j(k)}$ is the first moment of $\phi_{j(k)}$, and finally,
\begin{align}
    \mathcal{O}(\left|\braket{\phi_j|\phi_k}\right|^2)&= \left|\int\ud\w\phi_j^*(\w)\phi_k(\w)e^{i\w\tau_\text{S}}\right|^2 \nn \\
    &+\left|\braket{\phi_j|\phi_k}\right|^2\cos{\theta_{jk}}
\end{align}
are terms that depend on the overlap of $\phi_j$ and $\phi_k$ and are negligible except for when $\W_j\approx \W_k$, which is the regime where $p_{jk} \approx 0$. The main feature of $P_{jk}(\tau_\text{S})$ is the fringes from the oscillating term at the difference frequency $(\w_j-\w_k)$, which is a signature of frequency-bin entanglement \cite{Ramelow2009,Fedrizzi2009}, as these fringes arise due to the coherence between the two terms in the state $\ket{\Psi_{jk}}$.

In the absence of frequency resolution of the BSM, we again use $\hat\rho$ to obtain
\begin{align}
	P(\tau_\text{S}) &= \text{Tr}(\hat{\rho}\ \hat{\Pi}_\textrm{verif})\nn \\
	&= \sum_{jk} p_{jk}\int \ud\w\ud\w'\left|  \Braket{\w, \w'| \Psi_{jk}} \right|^2 \nn \\
	&= \sum_{jk} p_{jk} P_{jk}(\tau_\text{S}).
	\label{eq:fullptau:theory}
\end{align}
In terms of the previously defined quantities, we have
\begin{align}
	P(\tau_\text{S},\tau_\text{I})&=\frac{1}{4}
	\Bigg(
	1
	+\left|\int\ud\w\ud\W \, \left| f(\w,\W) \right|^2 e^{i(\w\tau_\text{S} + \W\tau_\text{I})}\right|^2 \nn \\
	&-\int \ud^2\W \, \modsqr{\rho_\mathrm{I}(\W,\W')}e^{i(\W-\W')\tau_\text{I}} \nn \\
	&- \int\ud^2\w \, \modsqr{\rho_\mathrm{S}(\w,\w')}e^{i(\w-\w')\tau_\text{S}}
		\Bigg),
	\label{eq:supp:peakfull}
\end{align}
%
%
where we have written the dependence of $\theta_{jk}$ on $\tau_\text{I}$ explicitly.

Eqs.\eqref{eq:supp:heraldedJSIsum} and \eqref{eq:fullptau:theory} show that summing over the individual ``pixellized" quantities requires a specific weight or gain. The results are naturally equivalent to replacing the spectrometers with bucket detectors. This process may also be interpreted in term of probabilities, where the probability of getting a four fold coincidence is the product of the probability of a coincidence between the heralding idler photons by the conditional probability of a coincidence between the heralded signal photons. This concept of weighting the average is similar to other experiments that utilize multipixel detection in the spatial \cite{treps2005} or spectral \cite{roslund2014,thiel2017} domain.

\section{Experiment}\label{sec:experiments}

\subsection{Description}
\begin{figure}[t]
    \centering
    \includegraphics[width=\linewidth]{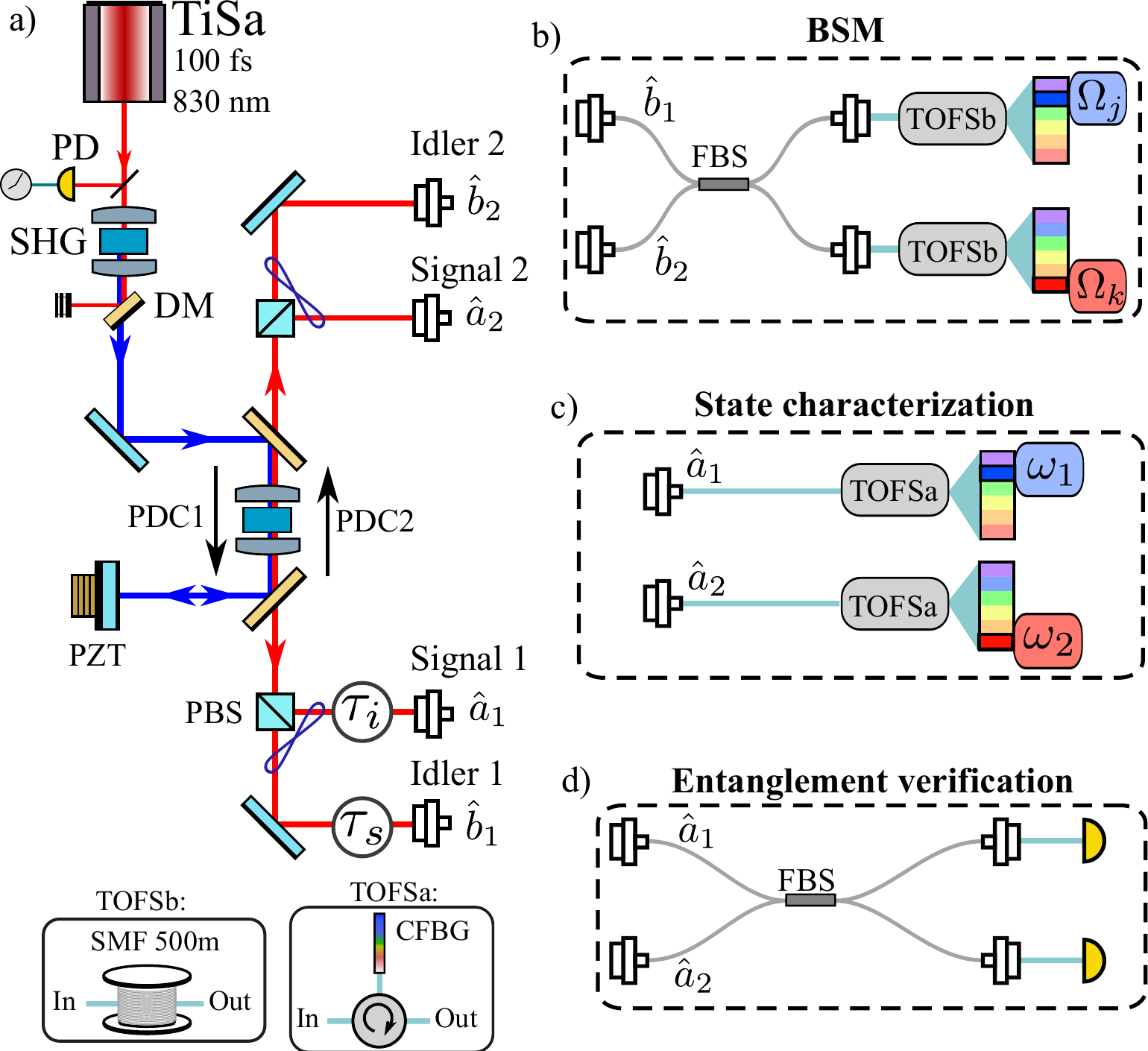}
    \caption{a) Experimental setup; PD: photodiode; SHG: second harmonic generation; DM: dichroic mirror; PDC: parametric down conversion; PZT: piezoelectric actuator; PBS: polarizing beamsplitter; TOFS: time of flight spectrometer; SMF: single mode fiber; CFBG: chirped fiber Bragg grating; FBS: fiber beamsplitter. b) Bell state measurement: the idler photons interferences are spectrally-resolved and used as a herald. c-d) different configuration of the measurement on the signal photons depending on the experiment.}
    \label{fig:exp:scheme}
\end{figure}
The BSM is achieved by interfering the idler photons from each source on a fiber beamsplitter which is used for heralding. Spectral resolution is then gained by utilizing frequency-to-time conversion at the output of the beamsplitter, thus heralding the frequency bins $\W_j$ and $\W_k$, see Fig.\ref{fig:exp:scheme}b). Finally, the heralded JSI and verification is done respectively by routing the signal photons through the setup described in Fig.\ref{fig:exp:scheme}c-d). An adjustable delay is introduced independently on the signal and idler paths of source 1 using delay lines to match the time of arrival of all the photons.

\subsection{Detection}
 
The single photon are detected utilizing superconducting nanowire single photon detector (SNSPD) from IDQuantique (ID281) which can detect the arrival time of photons with a resolution of 20 ps. This temporal resolution is translated into spectral resolution using time-of-flight spectrometers (TOFS), thanks to frequency-to-time conversion \cite{torres2011,goda2013}. For coarse spectral resolution, we used two  spools of 500 meters-long HP780 fiber. These imprint a dispersion of approximately 50 ps/nm and the losses per spool at 830 nm are less than 1 dB. For fine resolution, we used two chirped fiber Bragg gratings (CFBG from Teraxion) with a dispersion of 1000 ps/nm \cite{davis2017}. This extra resolution comes with heavy losses of over 10 dB from technical origins and from a finite spectral window of 10 nm. The photocurrent coming out of the detectors are registered with a time-to-digital converter (TDC, ID900 from IDQuantique). The time reference is provided by the clock generated by an internal photodiode in the laser source, thus ensuring that each time tag is taken with respect to a stable signal for each pulse. With this setup, it is possible to register coincidences between any combination of the four photons with spectral resolution. This allows for the ``pixelization'' of any event into spectral bins in post-processing.
\begin{figure}[t]
    \centering
    \includegraphics[width=.75\linewidth]{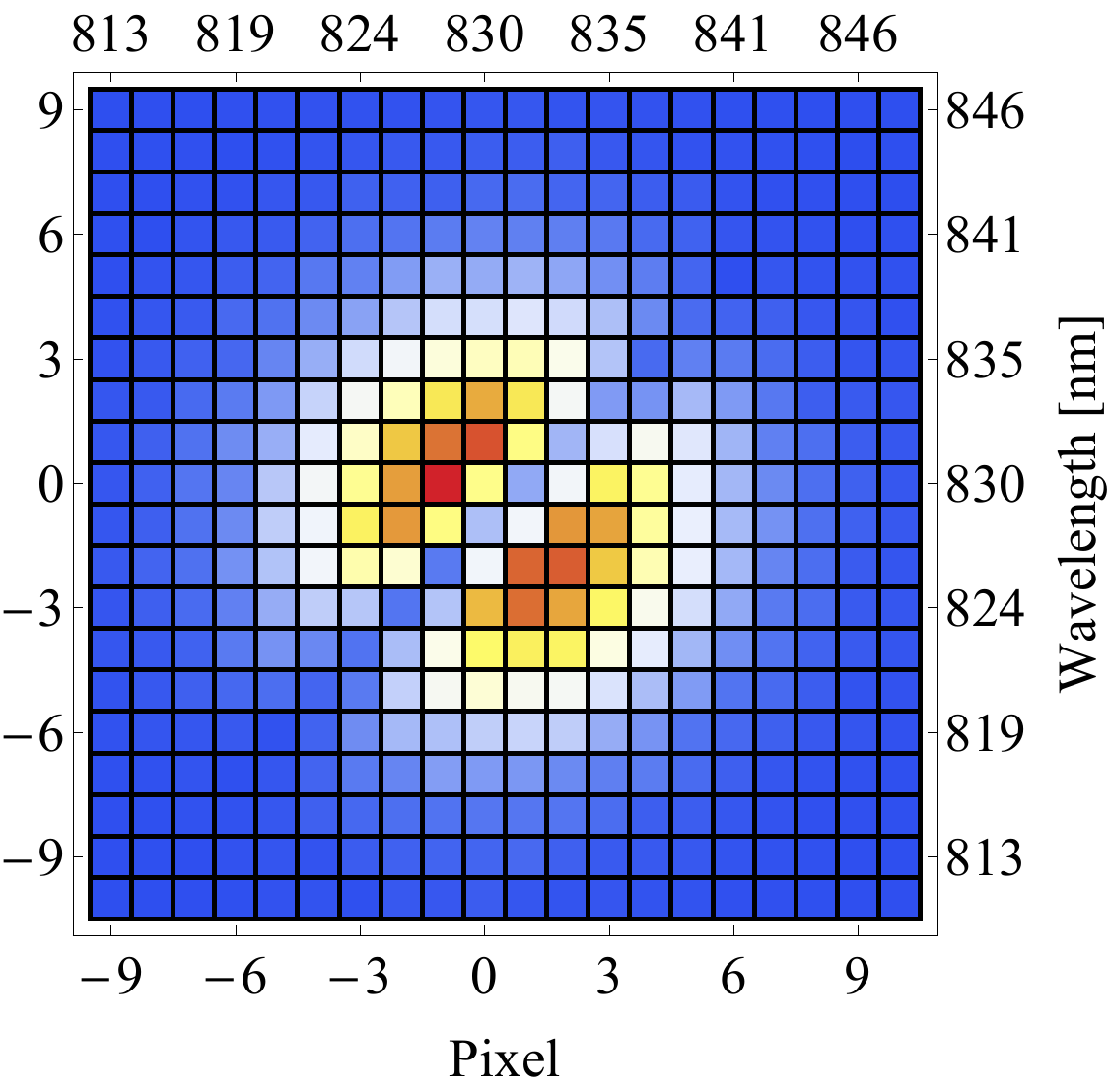}
    \caption{Acquisiton of the heralded JSI $p_{jk}$ of the idlers photon showing the calibration and indexing convention of the frequency bins.}
    \label{fig:pjk:exp}
\end{figure}

The data for the main experiments is acquired setting the time tagger's base unit at 100 ps, resulting in a spectral resolution of 0.1 nm when used with the CFBG and 2 nm with the fiber spools. For experiments that only require spectral resolution on the herald (such as the verification $P_{jk}$), time tags corresponding to heralding frequencies $\W_j,\W_k$ are acquired and subsequently histogrammed at the resolution of the spectrometer, corresponding to the probability of getting spectral coincidences in every possible combination of frequency bins. When all four spectrometers are needed (for instance to measure the JSI $F_{jk}$ heralded by a BSM at frequencies $\W_j$ and $\W_k$), the measurement then consists of four sets of time tags that can be binned into a four-dimensional histogram, such as shown in Fig.\ref{fig:supp:sim:Fjk}. For more readability, we label those heralding bins by integers $j$ and $k$, such that index $0$ corresponds to $\W_0$, the center wavelength of the idlers' joint spectrum. The axes in Fig.\ref{fig:pjk:exp} showcases the bin index (called pixel) to wavelength mapping.

\subsection{Source distinguishability}

This experiment relies critically on the indinstinguishability between the two sources as hinted in the previous section. This has to be achieved on every degree of freedom. Since the polarization and the spatial degree of freedom are constrained by polarization-maintaining fibers, there remains to match the sources in frequency and in time.

\begin{figure}[t]
	\centering
	\includegraphics[width=.85\linewidth]{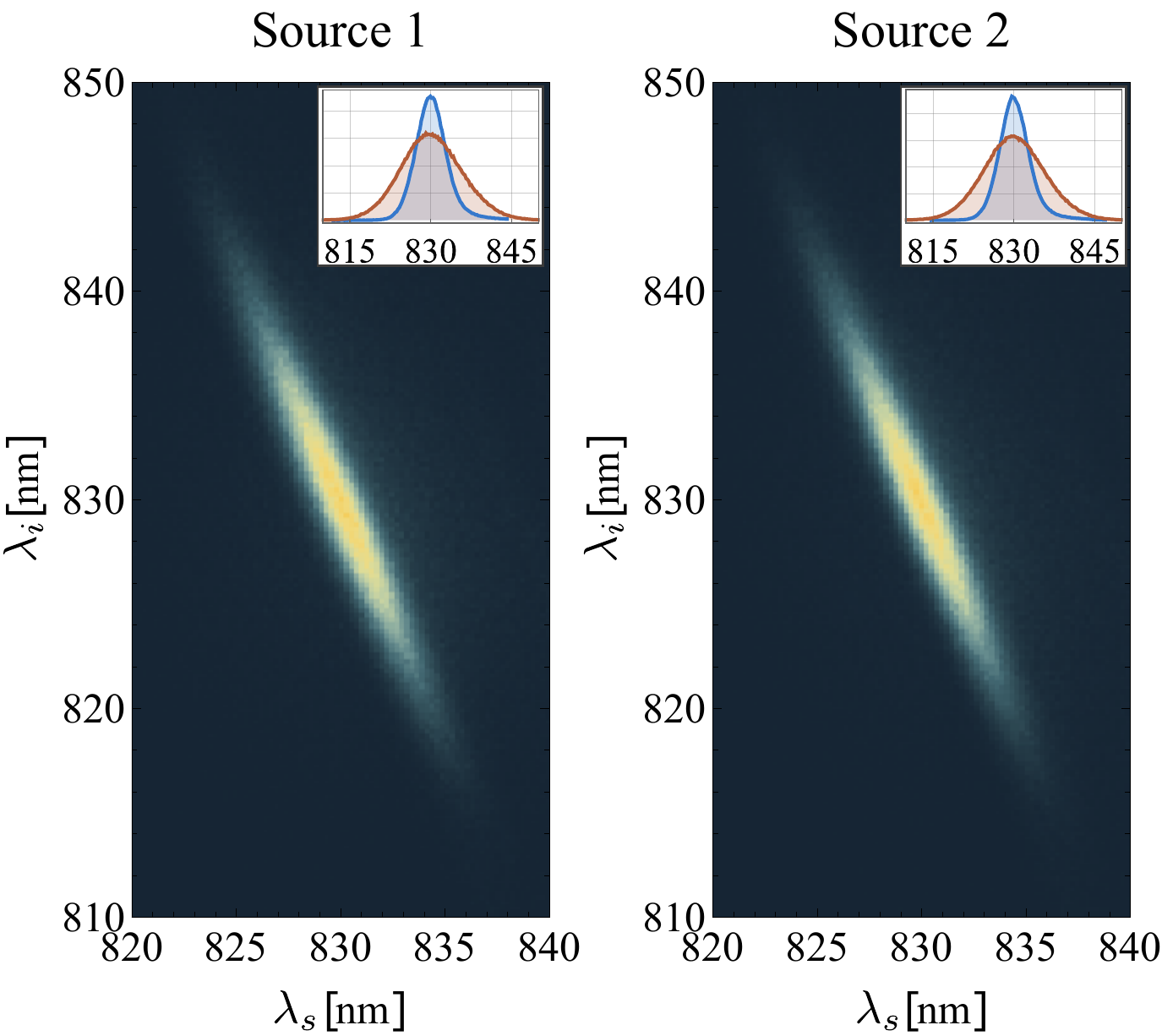}
	\caption{Experimental joint spectral intensity of both sources. Insets: marginal spectra.}
	\label{fig:supp:JSI}
\end{figure}

The spectral indistinguishability can be estimated by measuring the joint spectral intensity of both sources. This is steadily performed by connecting our spectrometers to the ports corresponding to $\adag_n$ and $\bdag_n$ in Fig.\ref{fig:exp:scheme}a), where $n=1,2$ is the source number, and acquiring the two-fold spectral coincidences. The JSIs, shown in Fig.\ref{fig:supp:JSI}, were obtained using the fiber spool as spectrometers but operating the time tagger at its maximum resolution of 13 ps, thus limiting spectral resolution by the timing jitter of the superconducting nanowires. We can see that both sources are very similar thanks to the dual pass configuration of the pump. This experimental JSI is used throughout this paper to define the parameters of the JSA $f(\w,\W)$ that are used in the simulation. Note that there can be an additional cause of distinguishability due to spectral phase mismatch between both sources which cannot be determined with an intensity measurement. Therefore, the dispersion was mostly matched in every path of the interferometer by ensuring that every fiber element had the same length. Any remaining dispersion would from free space propagation has a negligible effect.

The time indinstinguishability is achieved by scanning both delays $\tau_\text{S}$ and $\tau_\text{I}$ while monitoring the coincidences between ports $\hat c$ and $\hat d$ for the idlers and $\hat x$ and $\hat y$ for the signals, which statistics are uncorrelated. This results in a Hong Ou Mandel dip between the thermal states obtained by tracing over the other photon. The dips have a poor visibility (less than 10\%) which is still sufficient to coarsely match the time of arrival of the photons. Note that the same HOM interference was also realized by heralding, increasing the visibility. In Appendix \ref{app:purity}, we show how such measurement with and without spectral resolution can give a lower bound on the purity of the heralded state.

Finally, another convenient method \cite{kim2015} to match both sources can be applied. With the configuration from Fig.\ref{fig:supp:simple:scheme}, we can monitor the coincidences between either port of each beamsplitter, for instance $\hat{c}$ and $\hat{x}$, which share photon number correlations, as opposed to the previous uncorrelated case. When both delays are matched, i.e. $\tau_\text{S}=\tau_\text{I}=0$, then these coincidence oscillate at the optical frequency. This is akin to the classical case except that the interfering term involves a two-partite system. In Appendix \ref{app:ccfringes}, we show that the visibility of these interferences provides a direct measurement of the overlap between both sources, taking into account any phase effect. This method proved essential to accurately match delays before every experiment, while it also provided a bound to quantify indistinguishability, with a maximum measured contrast of $80\%$.

\section{Simulations and results} \label{sec:results}

In this section, we further model the experiment with a Gaussian approximation of the JSA to derive analytical expressions from the quantities defined in Sec. \ref{sec:theory}. This allows for a better understanding of the dependency of the interferences on the experimental parameters, notably on the delay between the idler photons in the BSM. We then compare our experimental results to the theory using this approximation. We first consider the case of a perfect BSM with $\tau_\text{I}=0$ and explore the effect of minor distinguishability in the BSM with $\tau_\text{I} \neq 0$.

\subsection{Gaussian model}

It is convenient to write the JSA $f(\w_\text{S},\w_\text{I})$ as a Gaussian distribution by approximating the Sinc function with a Gaussian of the same width :
\begin{align}
	f(\w_\text{S},\w_\text{I}) = C \, \exp \Bigg[ 
	&-\left(\frac{\w_\text{S}-\w_0}{2\sigma_\text{S}}\right)^2 	-\left(\frac{\w_\text{I}-\w_0}{2\sigma_\text{I}}\right)^2  \nonumber \\
	&-\alpha (\w_\text{S}-\w_0)(\w_\text{I}-\w_0)
	\Bigg], \label{eq:supp:JSAsource}
\end{align}
where $\sigma_{s}$ ($\sigma_\text{I}$) is the cross-sectional width of the JSI in the $\w_text{S}$- ($\w_\text{I}$-) direction evaluated at $\w_\text{I}$ ($\w_\text{S}$), $\w_0$ is the center frequency, $\alpha$ quantifies the amount of spectral entanglement, and $C=\left( \int \ud^2 \w \left| f(\w,\w') \right|^2 \right)^{-1/2}$ is a normalization constant. 

From this definition, the density matrices $\rho_\mathrm{S}$ and $\rho_\mathrm{I}$ are given by
\begin{equation}
\begin{gathered}
    \rho_\mathrm{S}(\w,\w')= C^2\sqrt{2\pi}\sigma_\text{I} \exp \Bigg[-\frac{(\w-\w_0)^2+(\w'-\w_0)^2}{4\sigma_\text{S}^2}\\
    +\frac{1}{2}\alpha^2\sigma_\text{I}^2(\w+\w'-2\w_0)^2\Bigg],
\end{gathered}
\end{equation}
\begin{equation}
\begin{gathered}
    \rho_\mathrm{I}(\W,\W')= C^2\sqrt{2\pi}\sigma_\text{S} \exp \Bigg[-\frac{(\W-\w_0)^2+(\W'-\w_0)^2}{4\sigma_\text{I}^2}\\
    +\frac{1}{2}\alpha^2\sigma_\text{S}^2(\W+\W'-2\w_0)^2\Bigg].
    \label{eq:rhoId}
\end{gathered}
\end{equation}
\begin{figure}[t]
	\centering
	\includegraphics[width=\linewidth]{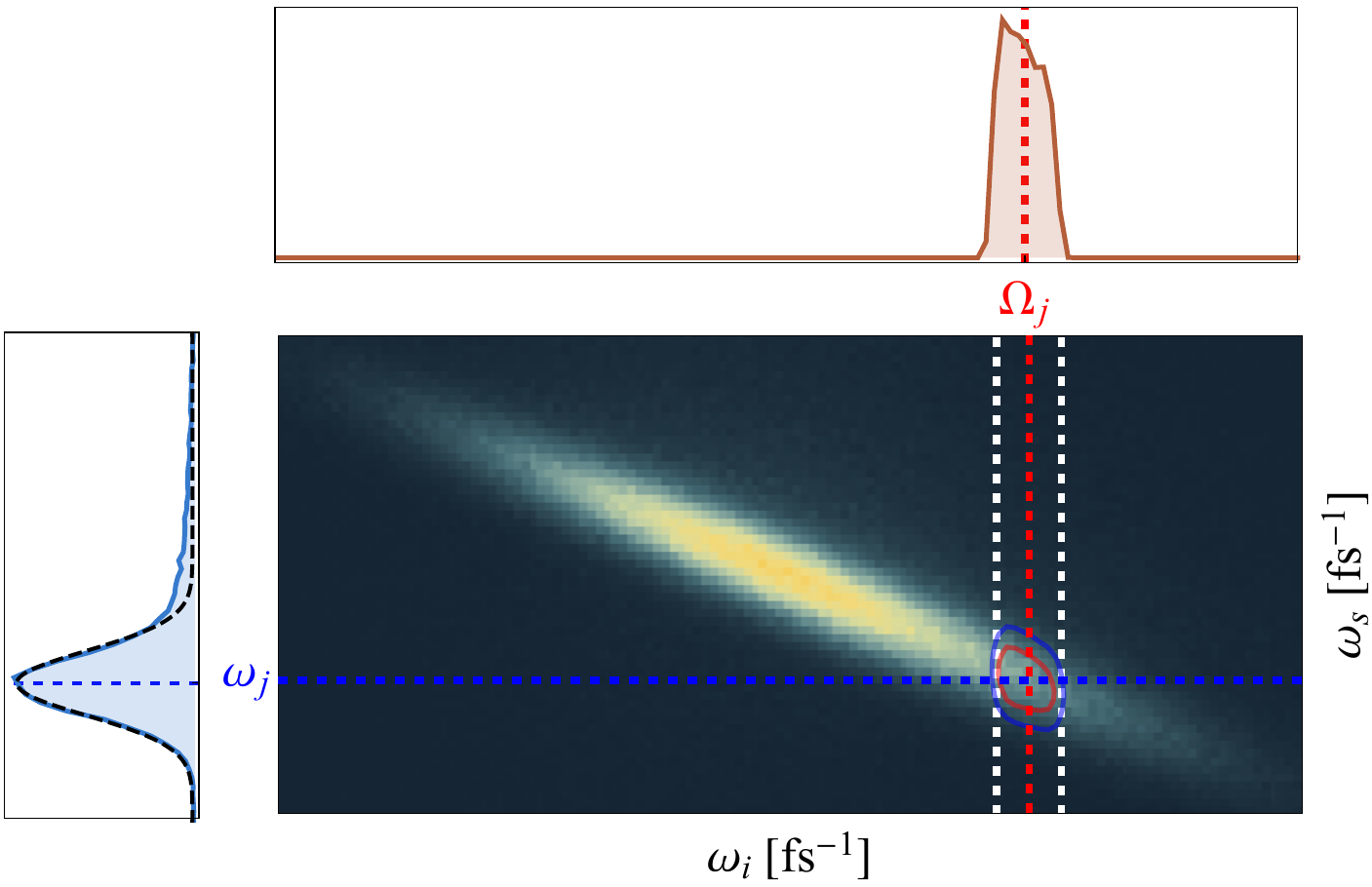}
	\caption{Gaussian approximation of the spectral heralding: a detection at frequency $\W_j$ projects the signal photon onto the mode $\phi_j$, centered at $\w_j$, shown on the left (dashed). The pure state model assumes that the idler photon is detected with perfect resolution and does not have any spectral support (top, dashed). A more accurate development shown in Appendix \ref{app:mixedstate} considers integration over a finite spectral window for the herald (top, plain) and the resulting heralded mode (left, plain). In this case, their first moment slightly differ from $\w_j$ and $\W_j$ obtained with the approximation.}
	\label{fig:mixedvspureJSI}
\end{figure}
Meanwhile, the $\phi_{j(k)}$ functions are given by
\begin{equation}
	\phi_{j(k)}(\omega) = \frac{1}{\sqrt{\sigma_\text{S}\sqrt{2\pi}}}\exp[-(\omega-\omega_{j(k)})^2/4\sigma_\text{S}^2],
	\label{eq:mode:phi:pure}
\end{equation}
which are Gaussians with a width equal to that of the signal's cross-sectional width and a central frequency given by
\begin{align}
	\omega_{j(k)}=\omega_0 - 2\alpha \ \sigma_\text{S}^2 \ \W_{j(k)}.
	\label{eq:supp:omegajk}
\end{align}
Finally, the normalization constant $\mathcal{C}_{jk}$ is given by
\begin{equation}
    \mathcal{C}_{jk} = 1-\exp\Big[-(\w_j-\w_k)^2/4\sigma_\text{S}^2\Big]\cos(\W_j-\W_k)\tau_\text{I}.
    \label{eq:Cjk}
\end{equation}
Note that, for most of our data, $(\w_j-\w_k)^2 \gg \sigma_\text{S}^2$, so that $\mathcal{C}_{jk}\simeq 1$. In Fig.\ref{fig:mixedvspureJSI}, we used the experimental JSI to simulate its marginals with a finite resolution. The heralded mode is centered at frequency $\w_j$ and we plotted in dashed lines $\modsqr{\phi_j}$ showing that the Gaussian approximation is quite accurate. We remind that we consider for the simulations that the spectral resolution along the idlers frequency is infinite.

\subsection{Case of $\tau_\text{I}=0$ ($\theta_{jk}=0)$}

We first study the case where the idler paths are exactly matched, setting $\tau_\text{I}=\theta_{jk}=0$, which yields the results that we report on in \cite{PRL2021}. Under this constraint, and using the Gaussian model in \eqref{eq:supp:JSAsource}, we proceed to obtain analytic forms for our measured quantities. First, the probability $p_{jk}$ of performing a BSM at frequencies $\W_j$ and $\W_k$ and heralding the state $\ket{\Psi_{jk}}$ is obtained from \eqref{eq:pjk:theory}:
\begin{align}
	p_{jk}=\frac{1}{2}\Big[ & \rho_\mathrm{I}(\Omega_j,\Omega_j)\rho_\mathrm{I}(\Omega_k,\Omega_k) - \modsqr{\rho_\mathrm{I}(\Omega_j,\Omega_k)} \Big].
	\label{eq:pjk:nodelay}
\end{align}
This in fact corresponds to the joint probability of detecting two photons at the output of a beamsplitter, when they are in a separable state at the input, with each photon described by a density matrix $\rho_\mathrm{I}$. Notably, the indistinguishability of the photons manifests as a dip along the degenerate $\W_j=\W_k$ frequencies as $p_{j=k}=0$. This implies that the probability of measuring any quantities is zero in this case. An experimental measurement of $p_{jk}$ is shown in Fig.\ref{fig:pjk:exp}, showing the bimodal structure of this distribution. It also shows our labelling convention for $\W_{j(k)}$ where we set $j(k)=0$ to correspond to $\W_0$ which is the center frequency of the down-converted light.

\begin{figure*}[t]
	\centering
	\includegraphics[width=1.16\linewidth]{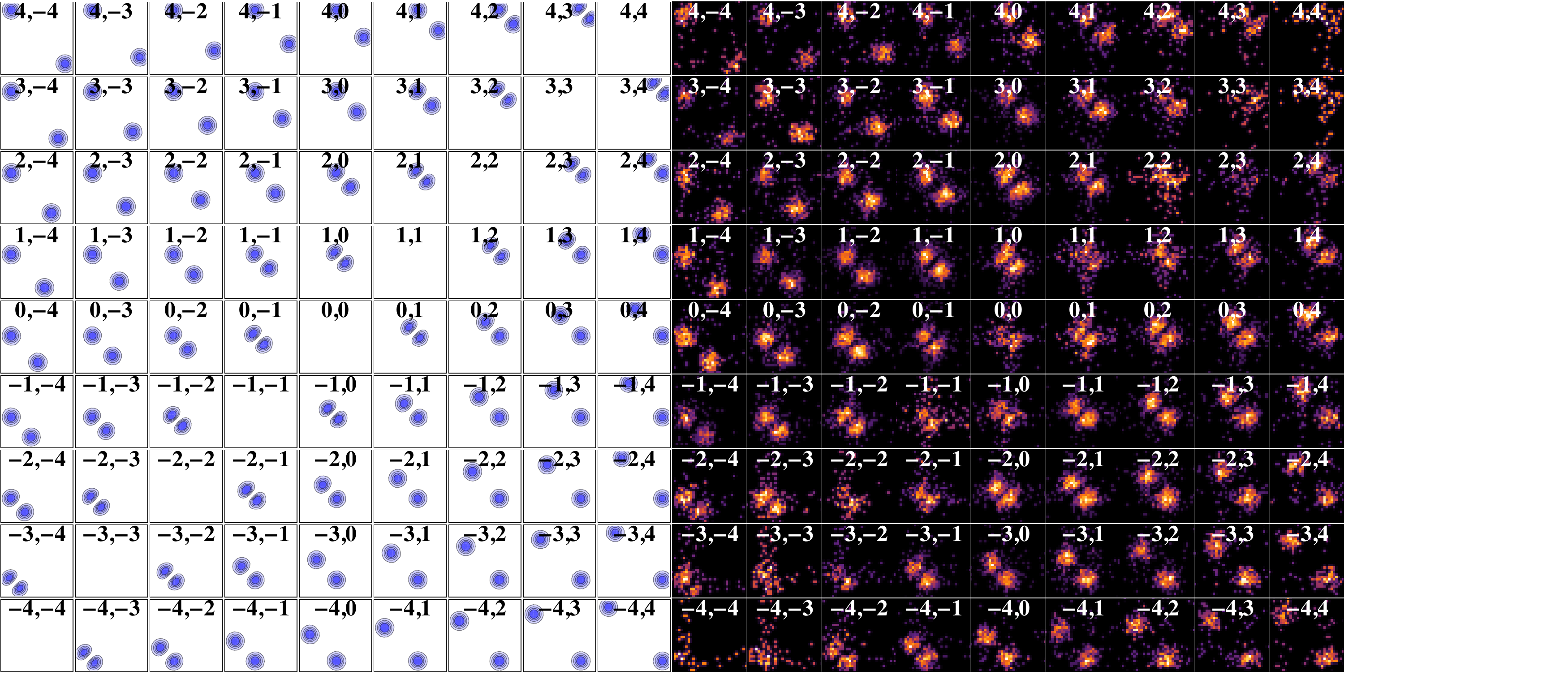}
    \caption{Left: simulation of the heralded JSI $F_{jk}$ for pure states, from Eq.\eqref{eq:heraldedJSI:nodelay1}, using experimental parameters derived from the sources' JSI measurement. Right: experimental result from \cite{PRL2021} obtained by acquiring $30 000$ spectral coincidences over 10 hours and binning the JSI into frequency bins, labelled according to Fig.\ref{fig:pjk:exp}.}
	\label{fig:supp:sim:Fjk}
\end{figure*}

Next we compute the JSI $F_{jk}$ associated with the state $\ket{\Psi_{jk}}$ from \eqref{eq:supp:heraldedJSI:delay}, which yields:
\begin{align}
    F_{jk}(\w_1,\w_2) = \frac{\left| \phi_j(\w_1)\phi_k(\w_2) -  \phi_j(\w_2)\phi_k(\w_1)\right|^2}{2 \mathcal{C}_{jk}}, \label{eq:heraldedJSI:nodelay1}
\end{align}
which can be approximated for $\left| \W_j-\W_k \right| \gg 0$, \textit{i.e.} for distant heralding frequencies, as
\begin{align}
    F_{jk}(\w_1,\w_2) \simeq  \frac{\left| \phi_j(\w_1)\phi_k(\w_2)\right|^2 + \left| \phi_j(\w_2)\phi_k(\w_1)\right|^2}{2}, \label{eq:heraldedJSI:nodelay}
\end{align}
which corresponds to two Gaussian spots centered at $(\w_j,\w_k)$ and $(\w_k,\w_j)$, hence mirror symmetric about the $\w_1=\w_2$ axis. For degenerate heralding events $\W_j=\W_k$, the heralded JSI is zero as the probability of performing a BSM in that case is null.

In the intermediate case where $\W_j \simeq \W_k$, the Gaussian spots are close to overlapped but the cross terms obtained by expanding \eqref{eq:heraldedJSI:nodelay1} maintains the symmetry of the heralded Bell state. These contribute to the JSI as a dip along the degenerate $\w_1=\w_2$ frequencies, akin to Eq.\eqref{eq:pjk:nodelay}. The resolved heralded JSI is simulated in Fig.\ref{fig:supp:sim:Fjk}a) for the pure state approximation, showing the aforementioned behaviour. The retrieve the same quantity experimentally, we set up the experiment for the characterization procedure (Fig.\ref{fig:exp:scheme}c) and measure spectral coincidences at the output of the signal FBS heralded by a BSM at frequencies $\W_j$,$\W_k$ (Fig.\ref{fig:exp:scheme}b). The result is shown in Fig.\ref{fig:supp:sim:Fjk}b) where the frequency bins are labelled according to Fig.\ref{fig:pjk:exp}. It is in good agreement with the pure state simulation, although the degenerate diagonal $j=k$ is not perfectly zero. This is due to the spectral resolution of the TOFS, where the degenerate frequency bins contain events corresponding to $\W_j\simeq\W_k$. Note also that both color maps are normalized, such that the amount of energy per bin has to be multiplied by the probability of realizing this measurement, given by $p_{jk}$ in Fig.\ref{fig:pjk:exp}. Therefore, experimentally, the amount of counts per bin exactly follows Fig.\ref{fig:pjk:exp}. In fact, the $p_{jk}$ distribution can be obtained by computing the total count of events detected per bin in Fig.\ref{fig:supp:sim:Fjk}b).

\begin{figure}[t]
	\centering
	\includegraphics[width=\linewidth]{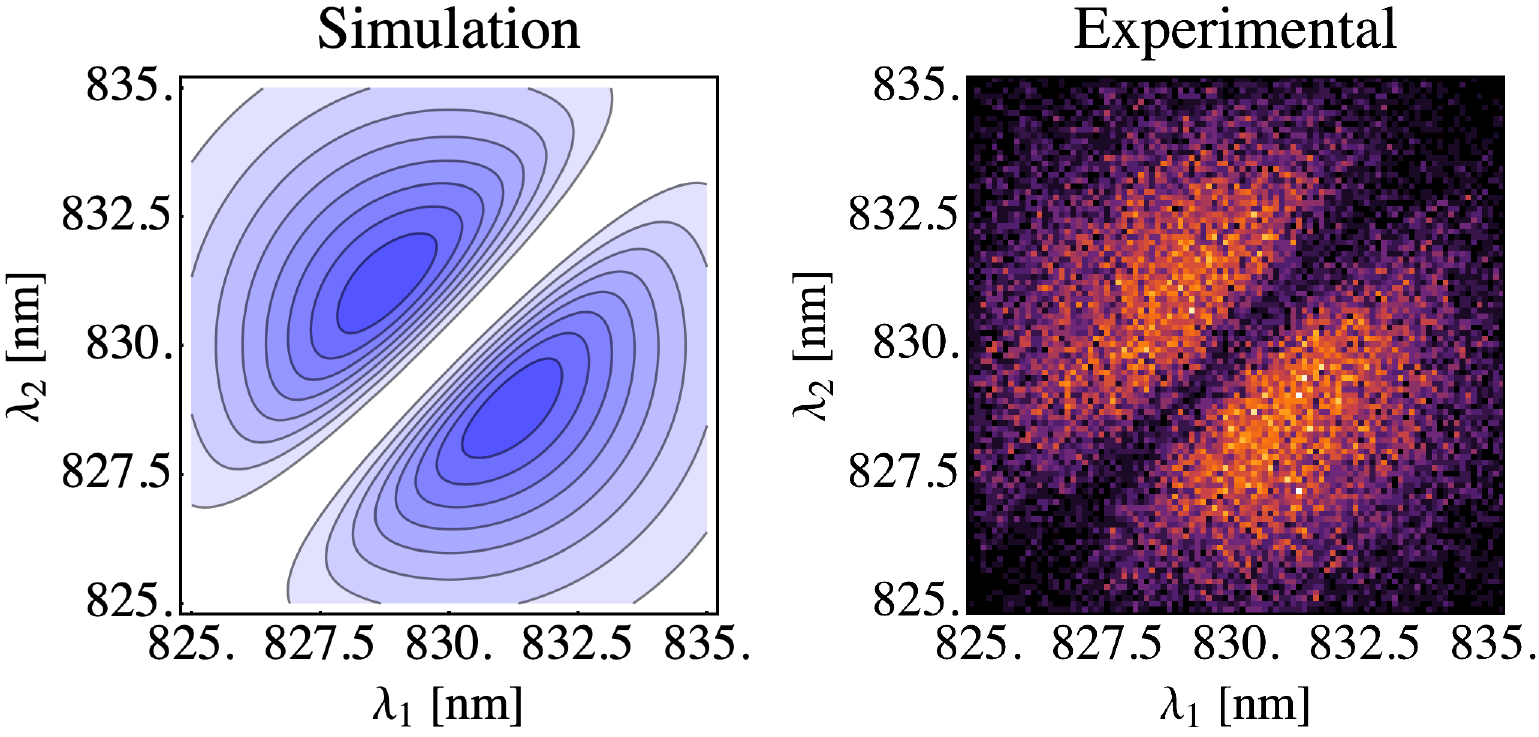}
    \caption{Left: simulation of the full heralded JSI $F$ defined by the sum of Eq.\eqref{eq:supp:heraldedJSIsum} over all $j,k$. Right: experimental result obtained by summing the acquisitions from Fig.\ref{fig:supp:sim:Fjk}b over all bins.}
	\label{fig:supp:sim:JSI}
\end{figure}

Tracing over the heralding frequencies, we obtain the mixed-state JSI $F(\w_1,\w_2)$ from Eq.\eqref{eq:heraldedJSI:full:delay}, which simplifies to
\begin{align}
    F(\w_1,\w_2) = \frac{1}{2}\left[\rho_\mathrm{S}(\w_1,\w_1)\rho_\mathrm{S}(\w_2,\w_2)-\left| \rho_\mathrm{S}(\w_1,\w_2) \right|^2\right],
    \label{eq:heraldedJSI:full:nodelay}
\end{align}
where we recall that we have multiplied each element $F_{jk}$ from Eq.\eqref{eq:heraldedJSI:nodelay1} by the proper weight $p_{jk}$ according to Eq.\eqref{eq:supp:heraldedJSIsum}. First we note that this is analogous to the idler JSI from Eq.\eqref{eq:pjk:nodelay}. Indeed, this is the joint spectral distribution that would be obtained were the beamsplitter placed in the signal paths rather than the idler paths, and the fact that such a distribution is measured without the presence of a beamsplitter is evidence of the non-local nature of this fourfold measurement. In Fig.\ref{fig:supp:sim:JSI}a), we simulate the heralded JSI with our approximated model, showing again the binodal structure characteristic of a Bell state.

With summing our experimental data from Fig.\ref{fig:supp:sim:Fjk}b over all heralded bins, we obtain the histogram shown in Fig.\ref{fig:supp:sim:JSI}b), which closely matches the simulation. We note that the JSI is indeed zero for degenerate frequencies, which is a consequence of photon-bunching. The measurements of the heralded JSI show the validity of the pure state approximation for the resolution of the heralding TOFS, which is sufficiently narrow.

We now proceed to evaluate the entanglement verification signal $P_{jk}(\tau_\text{S})$ from Eq.\eqref{eq:supp:PjkPure2}, for this case of $\theta_{jk}=0$. We find that $\mathcal{F}_j(\tau_\text{S})=\mathcal{F}_k(\tau_\text{S})$ since the Gaussians $\phi_{j(k)}$ only differ in their first moment $\w_{j(k)}$ (see Eqs.\eqref{eq:mode:phi:pure} and \eqref{eq:supp:omegajk}), such that we have:
\begin{align}
	P_{jk}(\tau) = &\frac{1}{2 \left( 1-|\braket{\phi_j|\phi_k}|^2 \right)} \times \nn \\
	&\Big\{ 1 + e^{-\sigma_\text{S}^2 \tau_\text{S}^2} \cos{\big[ (\w_j-\w_k) \tau_\text{S} \big]} \nn \\
	&-|\braket{\phi_j|\phi_k}|^2(1 + e^{-\sigma_\text{S}^2 \tau_\text{S}^2}) \Big\}.\label{eq:supp:PjkPure:nodelay}
\end{align}
Similar to the heralded JSI, we analyzing this function depending on the spectral distance between $\W_j$ and $\W_k$. We find that for $\left| \W_j - \W_k \right| \gg 0$, $\braket{\phi_j | \phi_k} \rightarrow 0$ such that the verification signal reduces to
\begin{align}
	P_{jk}(\tau_\text{S}) = \frac{1 + e^{-\sigma_\text{S}^2 \tau_\text{S}^2} \cos{\big[ (\w_j-\w_k) \tau_\text{S} \big]}}{2} .\label{eq:supp:PjkPure:nodelay:approx}
\end{align}
It corresponds to a background probability of $1/2$ and to a Gaussian envelope with a that corresponds to the transform limited temporal width of the signal's photon. This envelope is modulated by fringes at the difference of the heralded frequencies $\w_j-\w_k$. We note that the fringes vanish when setting $\alpha=0$ in \eqref{eq:supp:omegajk}, corroborating that the observation of fringes is a witness of entanglement swapping in the considered frequency bins.

\begin{figure*}[t]
	\centering
	\includegraphics[width=1.05\linewidth]{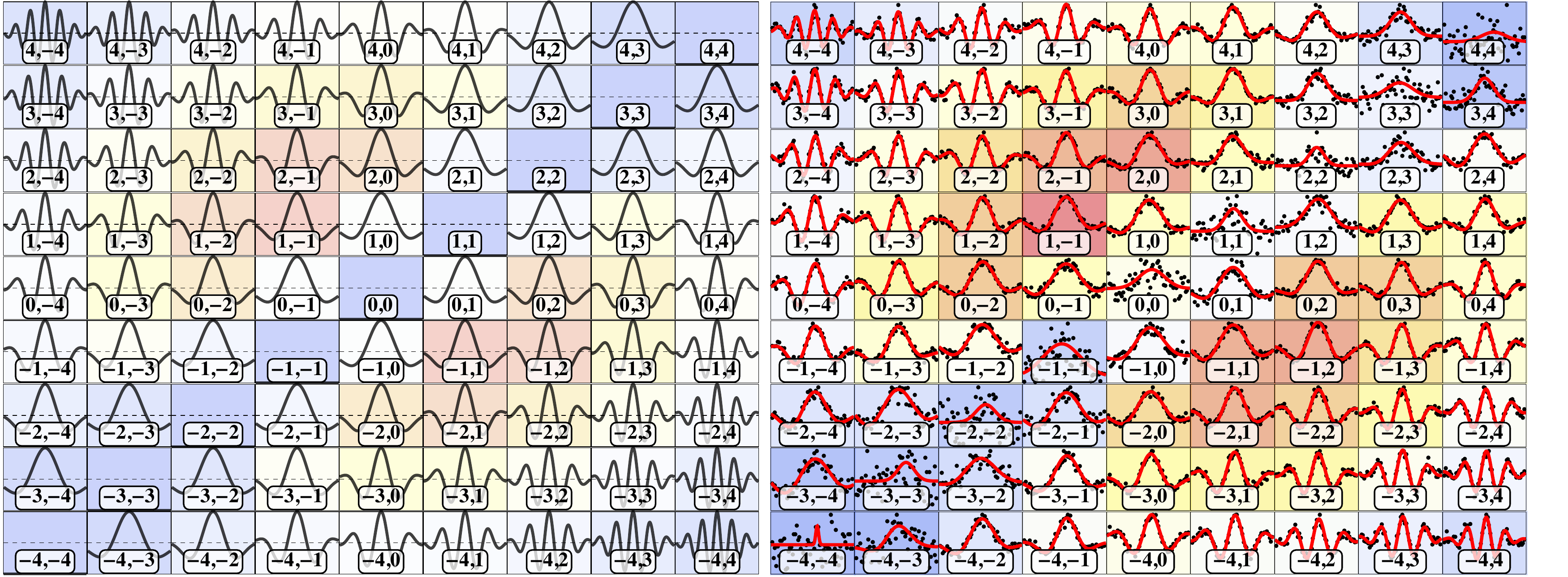}
    \caption{Left: simulation of the entanglement verification signal $P_{jk}$ for pure states, see Eq.\eqref{eq:supp:PjkPure:nodelay}. Right: experimental result from \cite{PRL2021}, obtained by binning the verification signal acquired over 15 hours. These results are fitted to the approximated model from \eqref{eq:supp:PjkPure:nodelay:approx}. The frequency bins are labelled according to Fig. \ref{fig:pjk:exp}. The colormap in the background represents the probability $p_{jk}$ of an heralding event, theoretical (top) and experimental (bottom).}
	\label{fig:supp:sim:Pjk}
\end{figure*}

Since in the pure state model, the probability $p_{jk}$ of obtaining a coincidence at degenerate frequencies $\W_j=\W_k$ is null, the verification signal \eqref{eq:supp:PjkPure:nodelay} is not defined so we set this $P_{j=k}=0$ for the simulation. However, we find that it has the following limit:
\begin{align}
    P_{j \rightarrow k}(\tau_\text{S}) \rightarrow \frac{1}{2} - \frac{1}{2}\left( 2\sigma_\text{S}^2 \tau_\text{S}^2 - 1 \right)e^{-\sigma_\text{S}^2 \tau_\text{S}^2} , \label{eq:supp:PjkPure:nodelay:approx2}
\end{align}
which can be easily demonstrated by noticing that $\braket{\phi_j | \phi_k}$ is a Gaussian of the variable $\w_j-\w_k$ under the Gaussian approximation (see for instance Eq.\eqref{eq:Cjk}). The expression is similar to that reported in \cite{Graffitti2020} that utilizes an engineered non-linearity to obtain spectral Bell states. Interestingly, the full signal $P(\tau_\text{S})$ that we describe next in Eq.\eqref{eq:Pfull:nodelay} has a similar expression with more complicated dependency on the experimental parameters and both describe the Gaussian peak sitting in a HOM dip. The former is a proof of entanglement whilst the latter results from HOM interferences for close-to-degenerate heralded frequencies. On Fig.\ref{fig:supp:sim:Pjk}a), we plotted the simulated $P_{jk}$ using our approximated model and parameters obtained experimentally. This plot shows the previously described behaviour, showing oscillations at the difference frequency which merge into a single peak in the near degenerate case. The background colormap for these plots represent the probability of measuring these events, \textit{i.e.} $p_{jk}$ from Eq.\eqref{eq:pjk:nodelay}.

In Fig.\ref{fig:supp:sim:Pjk}b), we show the experimental counterpart, with the colored background representing the experimental $p_{jk}$ obtained by computing the total number of counts in each bin. Again we see a nice agreement with the simulation except for the extreme bins which have a higher oscillating frequency in the simulated case. This is again the pure state approximation that results in $\phi_{j(k)}$ modes that are exactly centered at $\w_{j(k)}$ whereas in the experimental case, some averaging occurs due to the finite resolution of the TOFS. This is illustrated by Fig.\ref{fig:mixedvspureJSI} where the heralding and heralded center frequencies $\w_j$ and $\W_j$ can be different depending on the resolution of the spectral detection. Nevertheless, this shows that our approximated model, and most notably the limit derived in Eq.\eqref{eq:supp:PjkPure:nodelay:approx}, is sufficient to describe the entanglement swapping verification protocol, and we therefore use it to fit our results, shown as solid red curve in the experimental plot.

Finally we evaluate the verification signal $P(\tau_\text{S})$ with $\theta_{jk}=0$ for the non-spectrally-resolved case, given by Eq.\eqref{eq:fullptau:theory}. We find that it contains four terms:
\begin{align}
		P(\tau_\text{S})=\frac{1}{4}
		\Bigg[
		&1
		+\left|\int\ud\w\ud\W \, \modsqr{f(\w,\W)}e^{i \w\tau_\text{S}}\right|^2 \nn \\
		&-\int \ud^2\W \, \modsqr{\rho_\mathrm{I}(\W,\W')} \nn \\
		&- \int\ud^2\w \, \modsqr{\rho_\mathrm{S}(\w,\w')}e^{i(\w-\w')\tau_\text{S}}
		\Bigg].
		\label{eq:Pfull:nodelay}
\end{align}
The first one is simply a background probability, while the second one corresponds to the overlap integral between the two sources with a relative delay $\tau_\text{S}$ between the signal photons. Evaluating this term reveals a Gaussian along $\tau_\text{S}$ whose width depends on the joint temporal distribution of the sources. This is quite similar to the cross-correlation between two classical pulses, except that in the present case, the phase of the fringes is constant, implying that the $P_{jk}$ sum coherently to a single peak at $\tau_\text{S}=0$. We will see in the next section how that phase can be offset by introducing an additional time delay in the BSM.

\begin{figure}[t]
	\centering
	\includegraphics[width=\linewidth]{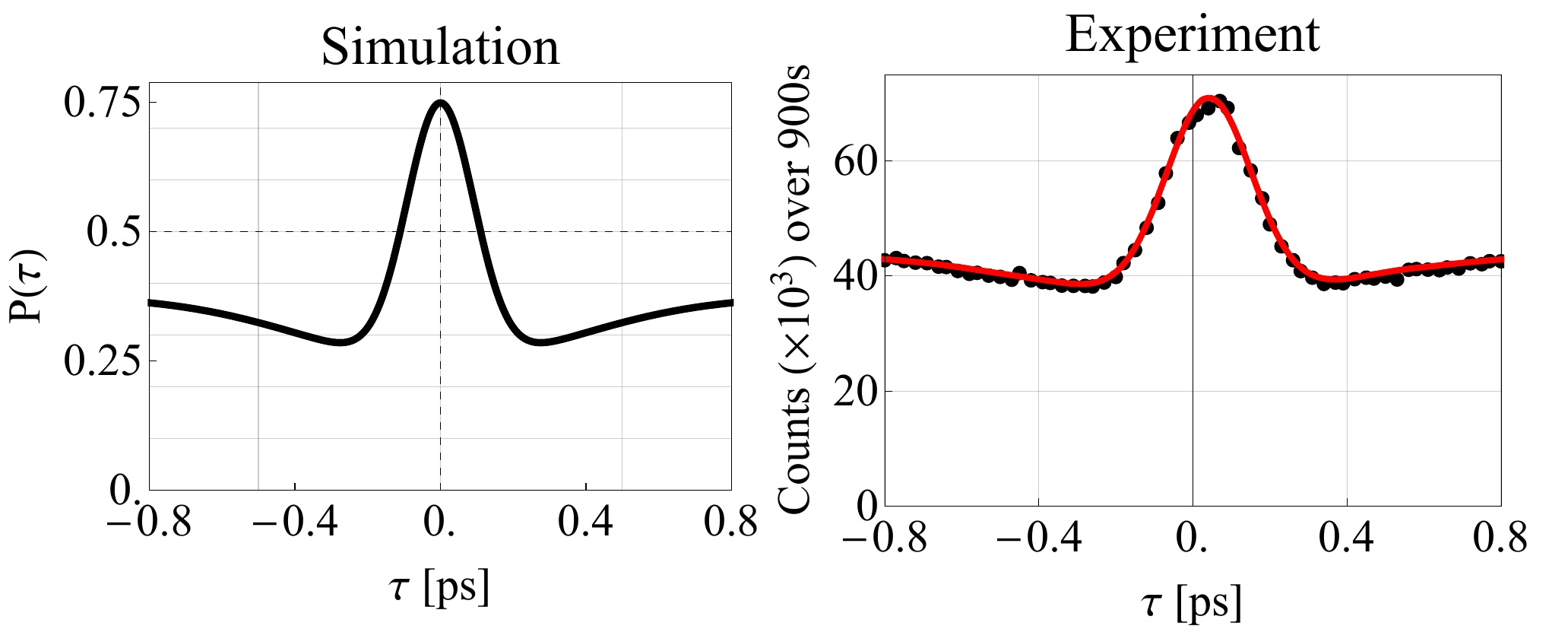}
	\caption{Probability $P(\tau)$ of a coincidence heralded by a BSM without spectral resolution. Left: simulation from Eq.\eqref{eq:Pfull:nodelay}; Right: experimental result presented in \cite{PRL2021} obtained by summing the individual $P_{jk}$ that are depicted in Fig.\ref{fig:supp:sim:Pjk}. The solid red curve is a sum of the individual fits on the experimental Gaussian model from Eq.\eqref{eq:supp:PjkPure:nodelay:approx}.}
	\label{fig:peak:theory:versus:experiment}
\end{figure}

The last two terms correspond respectively to the overlap integrals between the idler and the signal density matrices of each source. The former evaluate to a constant (which is unity when the sources are perfectly matched), while the latter describes an unheralded HOM dip between the signals photons. Hence, the full verification signal can be summarized as a Gaussian peak centered in a HOM dip. In Fig. \ref{fig:peak:theory:versus:experiment}a), we plotted a simulation of the full signal $P(\tau_\text{S})$ showing this behaviour. In Fig.  \ref{fig:peak:theory:versus:experiment}b), we show the experimental result obtained without binning that data from Fig.\ref{fig:supp:sim:Pjk}, which shows good agreement with the approximated model.

We stress that the presence of oscillating fringes in $P_{jk}$ or a peak in $P$, where the coincidence probability goes above the baseline of $1/2$, is a witness of an entangled state (see, for example, \cite{Fedrizzi2009}). Our setup is therefore capable of performing entanglement swapping between a large amount of frequency Bell states. As we show in \cite{PRL2021}, not all of the heralded states shown in Fig.\ref{fig:supp:sim:Fjk} are mutually orthogonal, but it is possible to select multiple subsets which form a set of mutually orthogonal Bell states. This is described in Appendix \ref{app:orthomodes}, where we run an algorithm to select 6 sets of 5 orthogonal modes, all of which satisfy the verification procedure for entanglement swapping.

\subsection{Case of $\tau_\text{I}\neq 0$ ($\theta_{jk}\neq 0$)}
\label{sec:results:delay}


We now study the case when there is a temporal delay $\tau_\text{I}$ between the idler photons. The resulting HOM interferences between them therefore have a decreased visibility. Note that when this delay becomes too large, the idler photons become distinguishable at the output of the beamsplitter, effectively removing the entanglement in the heralded Bell state. The formalism that we choose for our model necessitates to start from an entangled state. Therefore, in the scope of the paper, we will only consider a delay $\tau_\text{I}$ that is smaller than the inverse bandwidth $\sigma_\text{I}^{-1}\approx 300$ fs of the idler photons such that the state $\ket{\Psi_{jk}}$ retains some entanglement. One limitation of the pure state model arises because the idlers BSM \eqref{eq:supp:piBSMpure} is achieved with perfect resolution, leading to infinite temporal support of the quantities defined in Sec.\ref{sec:theory} over $\tau_\text{I}$ whereas they should necessarily be bounded by an envelope inversely proportional to the spectral resolution. In Fig.\ref{fig:mixedvspureJSI}, we see how the width of a realistic filtering of the idler photon differs from an infinitely narrow spectral detection. The derivation for the mixed state case is shown in Appendix \ref{app:mixedstate}. Nevertheless, the pure state model predict the correct physical behaviour and is therefore sufficient to describe the effects of slight distinguishability in the BSM.

The heralded state in the pure state approximation is given by Eq.\eqref{eq:supp:heraldedstate:delay} and shows a dephasing between the two states that depends on $\tau_\text{I}$. The heralded JSI from Eq.\eqref{eq:supp:heraldedJSI:delay} evaluates to:
\begin{align}
    F_{jk}(\w_1,\w_2) = \frac{1}{2 \mathcal{C}_{jk}}\Big[& \modsqr{\phi_j(\w_1)\phi_k(\w_2)} + \modsqr{\phi_k(\w_1)\phi_j(\w_2)} \nn \\
    &- 2\gamma_{jk}(\w_1,\w_2) \cos{(\W_j-\W_k)\tau_\text{I}} \Big],
\end{align}
where $\gamma(\w_1,\w_2) = \phi_j(\w_1)\phi_k(\w_2)\phi_j(\w_2)\phi_k(\w_1)$ for real modes $\phi_{j(k)}$, for simplicity. The last term is responsible for the HOM dip along the degenerate frequencies $\w_1=\w_2$, marking indistinguishability. As previously, $\gamma_{jk} \rightarrow 0$ when heralding distant bins $\left| \W_j - \W_k \right| \gg 0$ so the delay between the idlers photons has no influence. However, for degenerate bins $\W_j\rightarrow \W_k$ and for $\tau_\text{I} < \sigma_\text{I}^{-1}$, $\gamma_{jk}\times\cos\theta_{jk}\rightarrow 0$ which causes the two Gaussian spots to merge.

Putting these limits together, we find that a small delay between the idlers results in a heralded JSI similar to Fig.\ref{fig:supp:sim:Fjk}a) where the JSI in the bins close to the $j\rightarrow k$ diagonal are more or less merged depending on $\tau_\text{I}$. For spectral bins that are spaced further, the JSI is unchanged since the distinguishability is marked by a relative phase between the Gaussian spots.

The full JSI is obtained by summing $F_{jk}$ according to Eq.\eqref{eq:supp:heraldedJSIsum}: 
\begin{align}
	F(\w_1,\w_2) = \frac{\rho_\mathrm{S}(\w_1,\w_1) \rho_\mathrm{S}(\w_2,\w_2) 	- \Gamma(\w_1,\w_2;\tau_\text{I})}{2} ,
	\label{eq:heraldedJSI:full:delay2}
\end{align}
where $\Gamma$ is a function that depends on the overlap of the signal density matrices as a function of the idlers' delay (see Appendix \ref{app:math}). Similar to the previous case, the overlap between $\phi_j$ and $\phi_k$ depends on the indistinguishably in time of the idlers. 

We next consider the entanglement verification signal from Eq.\eqref{eq:supp:PjkPure2}, which depends on both $\tau_\text{S}$ and $\tau_\text{I}$ through $\theta_{jk}$. Evaluating it with the Gaussian model, we then obtain:
%
\begin{align}
	P_{jk}&(\tau_\text{S},\tau_\text{I}) = \frac{1}{2(1-|\braket{\phi_j|\phi_k}|^2\cos{\theta_{jk}})} \times \nn \\
	&\Bigg(1 + e^{-\sigma_\text{S}^2 \tau_\text{S}^2} \cos{\big[ \Delta\w_{jk} \tau_\text{S} - \theta_{jk} \big]} \nn \\ &-|\braket{\phi_j|\phi_k}|^2(e^{-\sigma_\text{S}^2 \tau_\text{S}^2} + \cos{\theta_{jk}})\Bigg). \label{eq:supp:Pjk:delay}
\end{align}
We perform yet another asymptotic behaviour analysis. For distant heralding frequencies, we find that the limit is similar to Eq.\eqref{eq:supp:PjkPure:nodelay:approx} with an additional dephasing:
\begin{align}
	P_{jk}(\tau_\text{S},\tau_\text{I}) = \frac{1 + e^{-\sigma_\text{S}^2 \tau_\text{S}^2} \cos{\big[ (\w_j-\w_k) (\tau_\text{S} - \tau_\text{I}') \big]}}{2} ,\label{eq:supp:PjkPure:delay:approx}
\end{align}
where we used \eqref{eq:supp:omegajk} to factorize by the difference of heralded frequencies and we defined $\tau_\text{I}' = \tau_\text{I} / 2\alpha \sigma_\text{S}^2$. We can see that the value of $\tau_\text{I}$ has a more noticeable effect compared to the heralded JSI. The fringes are no longer synchronized to the envelope and a phase shift occurs when the delay between the idlers photons is nonzero. This effect is quite important in our case since a delay as small as $100$ fs between the idler photons is sufficient to cause a dephasing of $\pi$, due to relatively large spectral bandwidth of the JSA.
\begin{figure}[t]
	\centering
	\includegraphics[width=\linewidth]{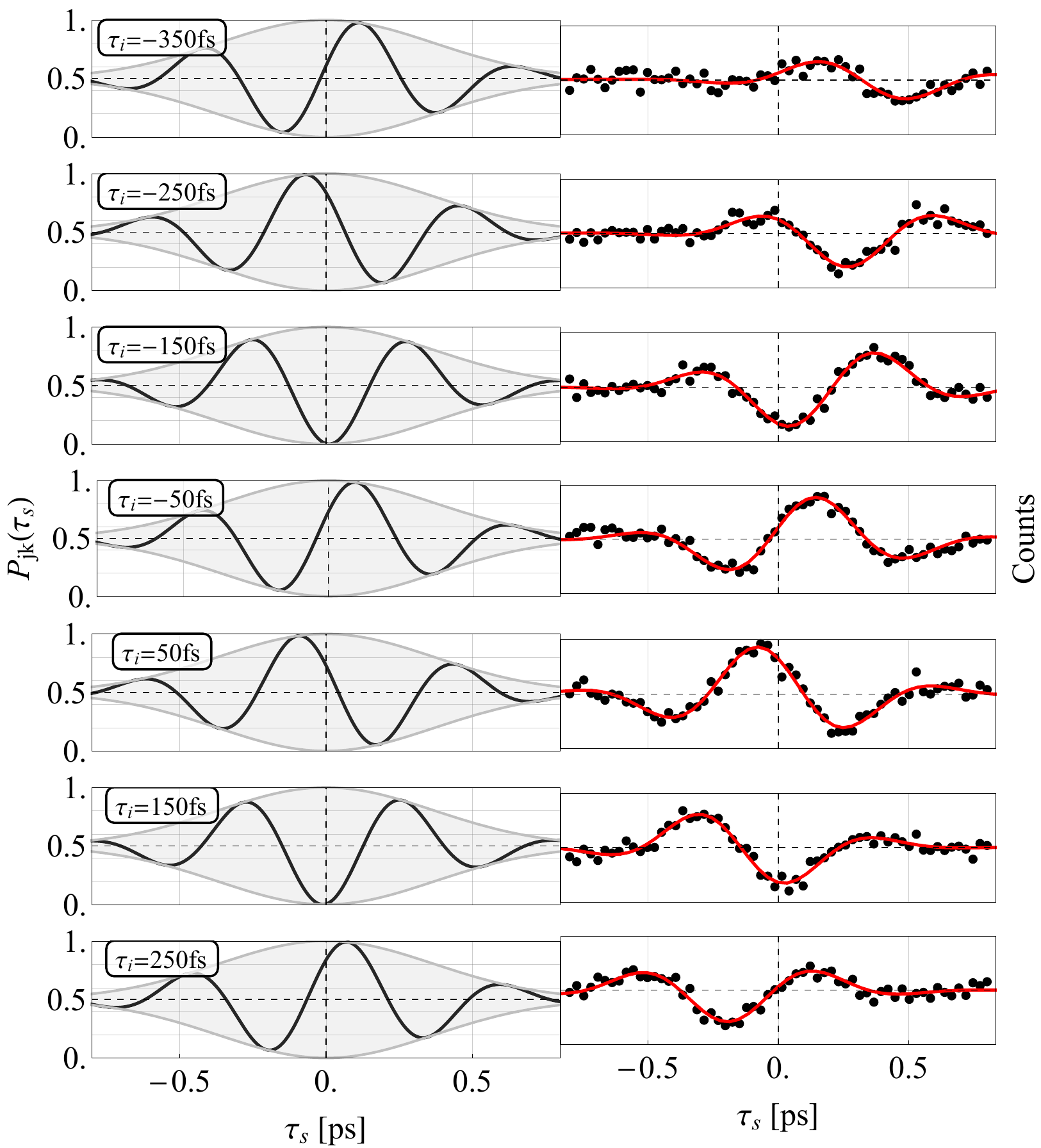}
	\caption{Simulated (left) and experimental (right) probability to get a coincidence heralded by a BSM at frequencies $\W_j$ and $\W_k$. The simulation utilizes the Gaussian, pure state approximation from Eq.\eqref{eq:supp:PjkPure:delay:approx} with the same parameters as in the experimental case. The experimental plots are acquired over 900s for different positions of the idler stage (on the right). The frequency bins used are separated by 8nm (or about 10 THz). The red curve represents a fit to the theoretical model. This is a different representation of the data shown in Fig.\ref{fig:exp:joydivision} in the bin labelled $(2,-2)$.}
	\label{fig:sim:fringesdelay}
\end{figure}
Note that using the more realistic mixed state model, the envelope of the fringes is also affected by this delay and causes a reduction in visibility. In the pure state approximation, the envelope in the idler's direction is infinite, but it is sufficient to show the most noticeable effect of the dephasing. Note also that it is possible to find a value for both delays such that $\tau_\text{S} = \tau_\text{I}'$, in which case there is no dephasing, but the visibility of the fringes would be decreased. This is illustrated in Fig.\ref{fig:sim:fringesdelay} which compares a simulation (left) that uses the approximated model with the same parameters as the experimental results (right). The entanglement verification utilizes the previous experimental protocol for 7 different values of $\tau_\text{I}$. While we see again very good agreement between the experimental and simulated fringes, the envelope in the model has no dependency on $\tau_\text{I}$ whereas it is clearly the case experimentally. 

For near-degenerate heralding frequencies, we have the following limit:
\begin{align}
	P_{j\rightarrow k}(\tau_\text{S},\tau_\text{I}) = \frac{1}{2} - \frac{1}{2}\cdot\frac{2 \sigma_\text{S}^2( \tau_\text{S} - \tau_\text{I}')^2 - 1}{1+ 4 (\tau_\text{I}')^2} e^{-\sigma_\text{S}^2 \tau_\text{S}^2} .\label{eq:supp:PjkPure:delay:approx2}
\end{align}
which is equal to Eq.\eqref{eq:supp:PjkPure:nodelay:approx2} when setting $\tau_\text{I}=0$. This represents again a Gaussian peak centered in a HOM dip at $\tau_\text{S}=0$. A change of variable delay between the idler photons then decreases its visibility.

Putting both limits together, we find that a scan over $\tau_\text{S}$ and $\tau_\text{I}$ of the verification signal $P_{jk}$ from Eq.\eqref{eq:supp:Pjk:delay} look very similar to Fig.\ref{fig:supp:sim:Pjk}a), except that the fringes will be offset as a function of $\tau_\text{I}$ while the peak close to the diagonal remains centered. In both cases, the visibility is decreased. This behaviour was very useful experimentally to verify that the delay at both FBS was as close to zero as possible. Moreover, the decrease in contrast for non-zero values of $\tau_\text{I}$ is yet another proof that the visibility of the oscillations in $P_{jk}$ is a marker of entanglement.
\begin{figure}[t]
	\centering
	\includegraphics[width=\linewidth]{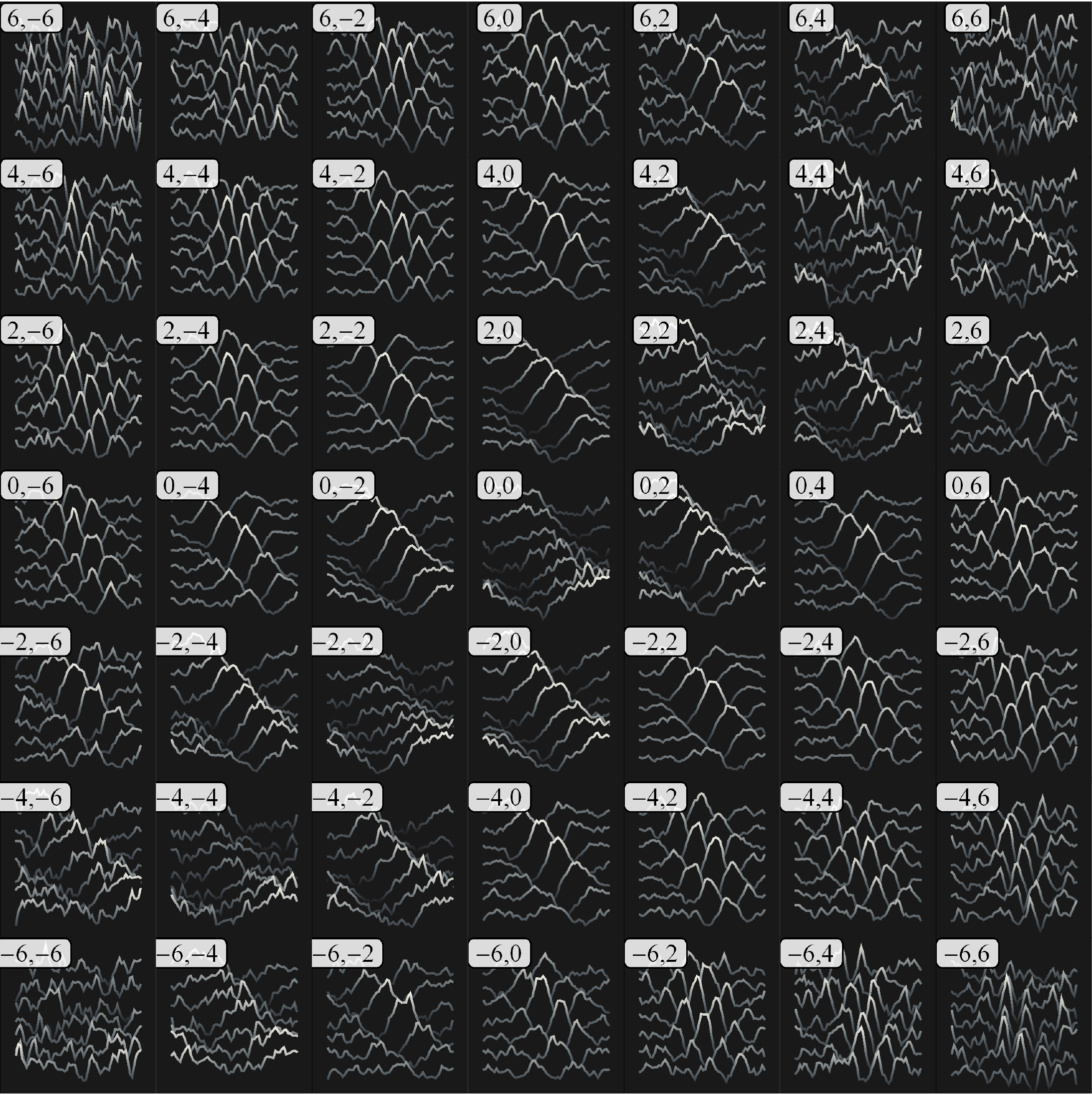}
	\caption{Waterfall plot of $P_{jk}$ plotted against $\tau_\text{S}$ (horizontal) for different values of $\tau_\text{I}$ (vertical) where each plot represents a frequency bin $\W_j,\W_k$ labelled by $j$ and $k$. The plot range and units for $\tau_\text{S}$ and $\tau_\text{I}$ are the same as shown in Fig. \ref{fig:sim:fringesdelay}}.
	\label{fig:exp:joydivision}
\end{figure}
In Fig.\ref{fig:exp:joydivision} we show an experimental waterfall plot representing each $P_{jk}$ for values of $\tau_\text{I}$ ranging from -300 to +300 fs which originates from the same dataset as Fig.\ref{fig:sim:fringesdelay}. Note that the spectrally resolved heralding is done at half the resolution than used earlier (shown in Fig.\ref{fig:pjk:exp}) to have sufficient statistics. These waterfall plots confirm the expected behaviour as we can see how the fringes are slanted in the $(\tau_\text{S},\tau_\text{I})$ space for distant $j,k$ frequency bins as predicted by \eqref{eq:supp:PjkPure:delay:approx}, while in the degenerate $j=k$ case, the fringes collapse to a single peak as described by Eq.\eqref{eq:supp:PjkPure:delay:approx2}. Note that the angle of the fringes is defined by the proportionality factor between $\tau_\text{I}$ and $\tau_\text{I}'$, which depends on the amount of entanglement and on the spectral bandwidth of the signal photons. These structures show that entanglement swapping is still achieved for this specific range of delay mismatch in the BSM, inducing distinguishability.

Finally, by repeating the same experiment either without spectral resolution of the herald or by summing the individual $P_{jk}$ according to Eq.\eqref{eq:fullptau:theory}, we find that the expression of the total verification probability is given by:
\begin{figure}[t]
	\centering
	\includegraphics[width=\linewidth]{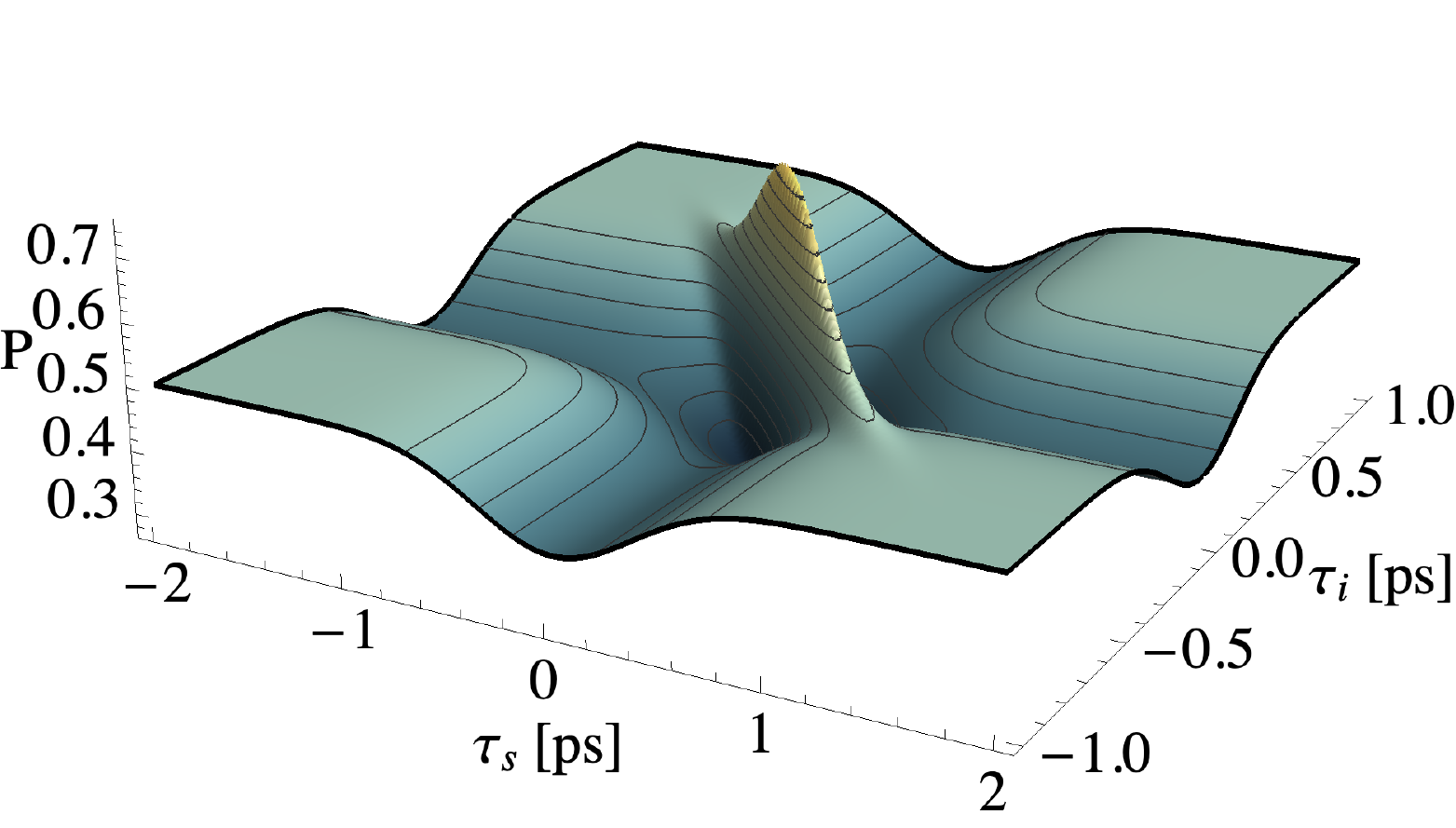}
	\includegraphics[width=\linewidth]{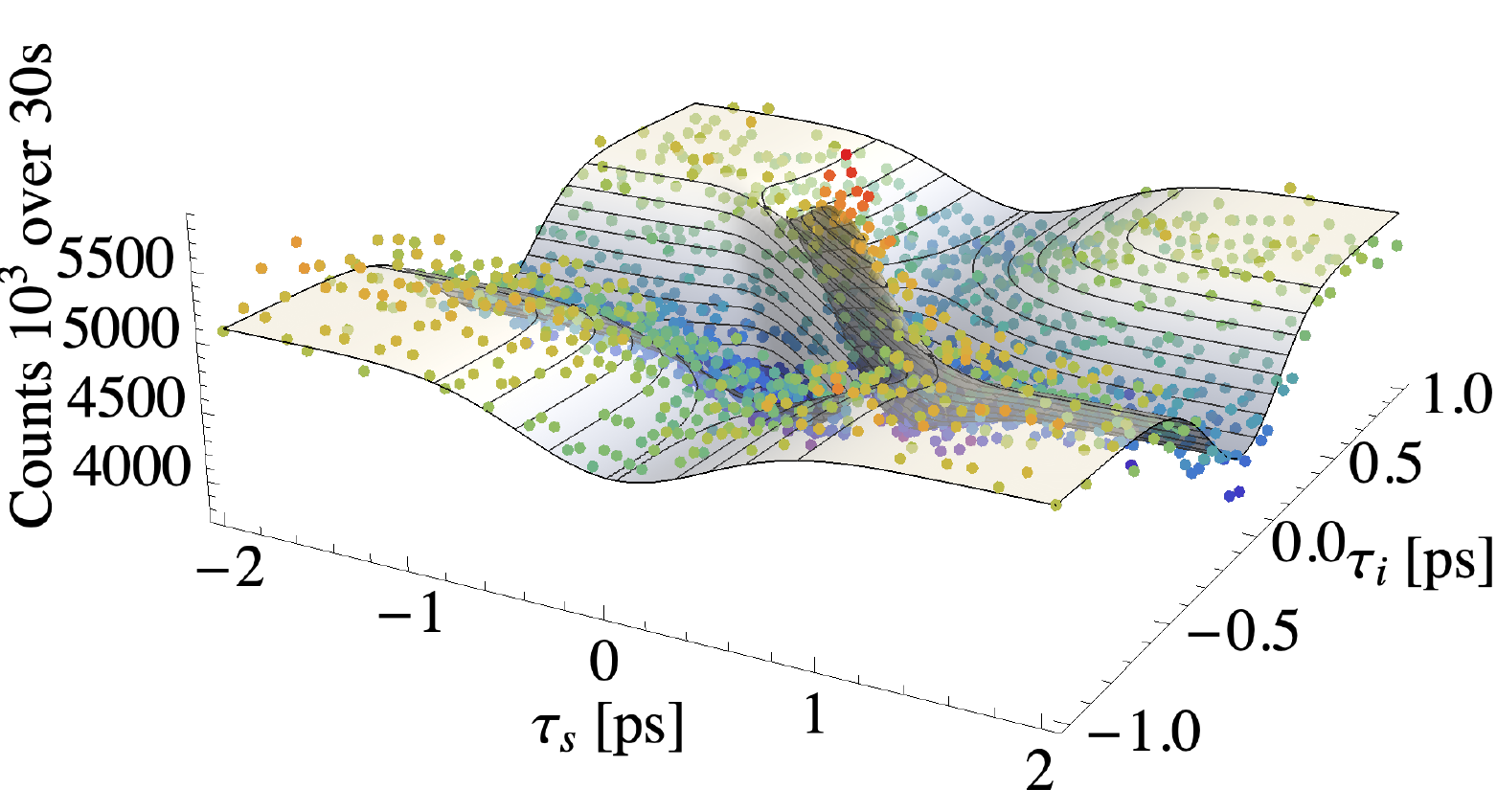}
	\caption{Top: Simulation of \eqref{eq:Pfull:delay} with experimental parameters. Bottom: experimental acquisition obtained by monitoring four-fold coincidences over 15 hours while scanning both $\tau_\text{S}$ and $\tau_\text{I}$ without spectral resolution of the herald, fitted to the function $P(\tau_\text{S},\tau_\text{I})$.}
	\label{fig:sim:2dpeak}
\end{figure}
\begin{align}
	P(\tau_\text{S},\tau_\text{I})=&\frac{1}{4}
	\Bigg(
	1
	+\left|\int\ud\w\ud\W \, \left|f(\w,\W)\right|^2 e^{i(\w\tau_\text{S} + \W\tau_\text{I})}\right|^2 \nn \\
	&-\int \ud^2\W \, \left| \rho_\mathrm{I}(\W,\W') \right|^2 e^{i(\W-\W')\tau_\text{I}} \nn \\
	&- \int\ud^2\w \, \left| \rho_\mathrm{S}(\W,\W') \right|^2 e^{i(\w-\w')\tau_\text{S}}
	\Bigg),
	\label{eq:Pfull:delay}
\end{align}
which is similar to Eq.\eqref{eq:Pfull:nodelay} with an additional dependency on $\tau_\text{I}$. We simulated this expression in Fig.\ref{fig:sim:2dpeak} and show an experimental acquisition as well. It is straightforward to identify the three nonconstant terms as familiar quantities. The second term is the cross-correlation between both JSA as a function of both delays. It can be written as the product of Fourier transform of the joint spectrum with respect to $\w$ and $\W$, thus reducing to the product of envelope functions centered at $\tau_\text{S}$ and $\tau_\text{I}$. This term is responsible for the slanted peak in Fig.\ref{fig:sim:2dpeak}.
The last two terms correspond respectively to the overlap integrals between the density matrices of the idlers and of the signals. They show the effect of interferences between uncorrelated photons and hence are visible as HOM dips along either $\tau_\text{S}$ or $\tau_\text{I}$, as shown again in Fig.\ref{fig:sim:2dpeak}. This figure allows us to identify more clearly the range over which entanglement swapping can be verified, which is essentially the area over which the slanted peak appears since the structure outside corresponds to quantum interferences between unentangled single photon states.

By exploring the range of both time delays in this configuration, we explicitly see the ultimate equivalence between signals and the idlers in what is effectively a four-photon interferometer. As such, it is adequate to consider $P_{jk}(\tau_\text{S},\tau_\text{I})$ and $P(\tau_\text{S},\tau_\text{I})$ as the most general representation of the phase-insensitive part of four-photon interference where time-frequency entanglement is present.

\section{Conclusion} \label{sec:conclusion}

In this work, we have undertaken a more thorough and more general analysis of the results we report on in Ref. \cite{PRL2021}. In summary, we demonstrated and analyzed a novel scheme for time-frequency entanglement swapping, using a multimode, spectrally-resolved Bell-state measurement as the heralding mechanism. The most salient feature of our method is the heralding of several, mutually-orthogonal Bell states derived from identical multimode entangled photon pairs. We further generalized our result to consider the case of non-zero time delay in the heralding Bell-state measurement, giving rise to Bell states with a varying amount of phase. Our setup is the first known to the authors to incorporate four simultaneous time-of-flight spectrometers, and thus points towards a promising venue of study of time-frequency entangled four-photon interferometry.
. 

\section*{Acknowledgements}
	This project has received funding from the European Union's Horizon 2020 research and innovation programme under Grant Agreement No. 665148, the United Kingdom Defense Science and Technology Laboratory (DSTL) under contract No. DSTLX-100092545, and the National Science Foundation under Grant No. 1620822.


\section*{Appendix}

\begin{appendix}

\section{Mixed state model} \label{app:mixedstate}

\subsection{Heralded state and JSI}

In the realistic case, the idlers BSM is not performed with perfect resolution, but rather with a finite spectral window. In our case, this is due to the resolution of the time-of-flight spectrometer, which is a convolution of multiple response functions in the frequency-to-time conversion. It is dominated by the timing jitter ($\simeq 20$ ps) of the superconducting nanowires.When this resolution is not perfect, then we can show that the signal photons are heralded into a mixed state. 

We begin by redefining the BSM operator as:
\begin{equation}
\begin{gathered}
	\hat{\Pi}_{lm}^\mathrm{BSM} =\int \ud\W_j \ud\W_k |t_l(\W_j)|^2  |t_m(\W_k)|^2\ \hat{\Pi}^{\text{BSM}}_{jk}\\
	=\int \ud\W_j \ud\W_k |t_l(\W_j)|^2  |t_m(\W_k)|^2 \ket{\W_j,\W_k}\bra{\W_j,\W_k},\\
	\ket{\W_j,\W_k}=\cdag(\W_j)\dddag(\W_k)\vac,
	\label{eq:pimixed}
\end{gathered}	
\end{equation}
where $t_{l(m)}(\W_{j(k)})$ are transmission amplitudes centered at $\W_{l(m)}$, and satisfy $\sum_{lm} |t_l(\W_j)|^2|t_m(\W_k)|^2 = 1$. We have introduced new indices $l$, $m$, so that we may incorporate the $j,k$-dependent quantities from the main text into our analysis. Moreover, it is straightforward to show that the POVM \eqref{eq:supp:piBSMpure} is obtained by setting $t_{l(m)}(\W) \rightarrow \delta (\W-\W_{l(m)})$. To simplify our notation henceforth, we will use the following shorthand
\begin{equation}
    \iint_{lm} \ud \W_j \ud \W_k := \int \ud\W_j \ud\W_k |t_l(\W_j)|^2  |t_m(\W_k)|^2.
\end{equation}
It can be seen from Fig.\ref{fig:mixedvspureJSI} that the finite resolution of the idler detection bin reduces the purity of the heralded signal state, due to the entanglement of the two-photon state. Because of this, the POVM element \eqref{eq:pimixed} requires that we describe the heralded state as a mixed state
\begin{align}
	\hat{\rho}_{lm} = \frac{\operatorname{Tr}_{\hat{b}} \Big[ \hat{\Pi}_{lm}^\mathrm{BSM} \ket{\psi_{12}} \bra{\psi_{12}} \Big]}{p_{lm}},
\end{align}
where $\operatorname{Tr}_{\hat{b}}$ is the partial trace over the subspace defined by operators $\hat{b}_1$ and $\hat{b}_2$. Analogously to the pure state case, the probability $p_{lm}$ are defined as
\begin{align}
	p_{lm} &= \operatorname{Tr} \Big[ \hat{\Pi}_{lm}^\mathrm{BSM} \ket{\psi_{12}} \bra{\psi_{12}} \Big] \nonumber\\
	&= \frac{1}{2} \iint_{lm} \ud \W_j \ud \W_k \nonumber \Big( \rho_\mathrm{I}(\W_j,\W_j) \rho_\mathrm{I}(\W_k,\W_k) \nonumber\\
	&\quad -  \rho_\mathrm{I}(\W_j,\W_k) \rho_\mathrm{I}(\W_k,\W_j) \Big) \nonumber \\
	&= \iint_{lm} \ud \W_j \ud \W_k\ p_{jk}.
\end{align}
where the idler density matrix is defined as Eq.\eqref{eq:rhoz}, and we again obtain Eq.\eqref{eq:pjk:theory} by setting the filters $t_{l(m)}$ as $\delta$ functions.

We may now compute the heralded state density matrix:
\begin{equation}
	\hat{\rho}_{lm} = \frac{1}{p_{lm}}\iint_{lm} \ud \W_j \ud \W_k\ p_{jk}\  \ket{\Psi_{jk}}\bra{\Psi_{jk}}
	\label{eq:rholm}
\end{equation}
while recalling the definition of $\phi_{j(k)}(\w)$ in $\ket{\Psi_{jk}}$ as
\begin{equation}
    \phi_{j(k)}(\w) = \frac{f(\w,\W_{j(k)})}{\sqrt{\rho_\mathrm{I}(\W_{j(k)},\W_{j(k)})}}.
\end{equation}
This has the intuitive interpretation of a mixed state as an incoherent sum of pure states over the detection bandwidth of $t_{l(m)}$. All of the measured quantities follow in a straightforward manner. In the absence of frequency resolution, we herald again the mixed state
\begin{equation}
    \hat{\rho}=\sum_{lm}\ p_{lm}\ \hat{\rho}_{lm},
\end{equation}
just as with the pure state model.

The heralded JSI is given by:
\begin{align}
	F_{lm}(\w_1,\w_2) = \bra{\w_1,\w_2} \hat{\rho}_{lm} \ket{\w_1,\w_2},
\end{align}
which can be expressed in terms of $F_{jk}$ as
\begin{align}
    F_{lm}(\w_1,\w_2) = \frac{1}{p_{lm}}\iint_{lm} \ud \W_j \ud \W_k\  p_{jk}\ F_{jk}(\w_1,\w_2).
\end{align}
The integrated JSI corresponding to the state $\hat{\rho}$ is again given by
\begin{align}
	F(\w_1,\w_2)= \sum_{lm} p_{lm} F_{lm}(\w_1,\w_2).
\end{align}

When the signal photons in the state $\hat{\rho}_{lm}$  are incident on a 50:50 beamsplitter, the coincidence fringes at the output are given by
\begin{align}
	P_{lm}(\tau_\text{S},\tau_\text{I}) = \text{Tr} \left(\hat{\Pi}_\text{verif} \, \hat{\rho}_{lm}\right)=\int \ud^2\w\bra{\w,\w'}\hat{\rho}_{lm}\ket{\w,\w'},
\end{align}
where $\ket{\w,\w'}=\xdag(\w)\ydag(\w')\vac$ as before. Evaluating this in terms of $P_{jk}$, we 
\begin{align}
	P_{lm}(\tau_\text{S},\tau_\text{I}) = \frac{1}{p_{lm}} \iint_{lm} \ud\W_j \ud\W_k\  p_{jk}\ P_{jk}(\tau_\text{S},\tau_\text{I}).
\end{align}
Finally the integrated interference peak for the state $\hat{\rho}$ is recovered by taking
\begin{equation}
    P(\tau_\text{S},\tau_\text{I}) = \sum_{lm}\ p_{lm}\ P_{lm}(\tau_\text{S},\tau_\text{I}).
\end{equation}

We will make a few comments regarding the most interesting feature of comparing this model with the pure state model. To this end, we shall represent the pure state $\ket{\Psi_{jk}}\bra{\Psi_{jk}}$ as a density matrix in the $\{\ket{\phi_j}\ket{\phi_k},\ket{\phi_k}\ket{\phi_j}\}$ basis as follows:
\begin{equation}
    \ket{\Psi_{jk}}\bra{\Psi_{jk}} = \frac{1}{2\mathcal{C}_{jk}}
                                    \begin{pmatrix}
                                     1 & e^{i\theta_{jk}}\\
                                    -e^{-i\theta_{jk}} & 1
                                    \end{pmatrix}.
\end{equation}
Meanwhile, the mixed state $\hat{\rho}_{lm}$ has the representation:
\begin{equation}
    \hat{\rho}_{lm} = \frac{1}{p_{lm}}\iint_{lm} \ud \W_j \ud \W_k\ \frac{p_{jk}}{2\mathcal{C}_{jk}}\ 
                                    \begin{pmatrix}
                                     1 & e^{i\theta_{jk}}\\
                                    -e^{-i\theta_{jk}} & 1
                                    \end{pmatrix}.
\end{equation}

Note that the off-diagonal terms $e^{\pm i \theta_{jk}} = e^{\pm i (\W_j-\W_k)\tau_\text{I}}$, the coherence terms, are the hallmark of the bipartite entanglement in these states, and are responsible for the interference we observe in $P_{jk}$. It seems reasonable then to ask if, and to what extent, the averaging over the $l,m$ bands in the mixed state is expected to reduce the entanglement. Take first the special case of $\tau_\text{I} = 0$. In this case, $\theta_{jk} = 0\ \forall\ j,k$, and the off-diagonal terms are equal to unity, and hence, no phase-averaging occurs for the mixed state $\hat{\rho}_{lm}$. Indeed, the off-diagonals survive even for the fully-averaged state $\hat{\rho}$, and one can interpret this as the reason why the interference peak survives at full visibility for $\tau_\text{I}=0$.

More generally, however, for $\tau_\text{I} \neq 0$, the disagreement between the two models becomes more salient. The pure-state model assumes that the idler photons are detected at monochromatic frequency modes at $(\W_j, \W_k)$, which are necessarily infinite in extent in the time domain. This means that the idler photons remain indistinguishable at the output of the BSM beamsplitter, even for arbitrary $\tau_\text{I}$ delays at the input. On the other hand, taking into account the averaging over the $l,m$ bands introduces distinguishability, and the BSM is no longer ideal in this case. Furthermore, the phase-averaging becomes worse the larger $\tau_\text{I}$ is, since the argument of the phase factor is $(\W_j-\W_k)\tau_\text{I}$. This feature results in a reduction of the off-diagonal terms, and of the visibility of the $P_{lm}$ interference for large enough $\tau_\text{I}$. This behavior is most clearly seen in the profile of the fully-integrated two-dimensional peak, which results from averaging over all phases with a weight $p_{jk}$. The peak vanishes in the $\tau_\text{I}$ direction over a delay timescale comparable to the inverse bandwidth of the idler photons. The intuitive interpretation is that the idler photons become distinguishable at the output of the beamsplitter when the relative delay is at least as long as their pulse durations.

\section{Source distinguishability} \label{app:ccfringes}

The entanglement verification protocol we use, that is, the two-photon interference of the state $\ket{\Psi_{jk}}$, ultimately relies on the indistinguishability of the two source states. To see this, we relabel the source JSA's as $f_1(\w,\W)$ and $f_2(\w,\W)$, and for simplicity, we assume that they are identical up to a translation in frequency space. Note now that this leads to a heralded state

\begin{align}
	\ket{\Psi_{jk}} \propto \ket{\phi^1_j}_1\ket{\phi^2_k}_2-\ket{\phi^1_k}_1\ket{\phi^2_j}_2,
\end{align}
where
\begin{align}
	\ket{\phi^{1(2)}_{j(k)}} = \int \ud \w \phi^{1(2)}_{j(k)}(\w)\adag_{1(2)}(\w)\vac,
\end{align}
and
\begin{align}
	\phi^{1(2)}_{j(k)}(\w) = \frac{f_{1(2)}(\w,\W_{j(k)})}{\rho_{1(2)}(\W_{j(k)},\W_{j(k)})}.
\end{align}

Although this state is still entangled, the verification method using coincidence fringes in $P_{jk}(\tau)$ will suffer from a reduction in visibility due to the distinguishability of $f_1$ and $f_2$. To see this, we recalculate $P_{jk}(\tau)$ in its approximate form \eqref{eq:supp:PjkPure:delay:approx}, and find

\begin{align}
	P_{jk}(\tau_S) \approx \frac{1}{2}\left(1-V_{jk} \, e^{\sigma_\text{S}^2\tau_S^2}\cos{\left[\left(\frac{\w^1_j+\w^2_j}{2}-\frac{\w^1_k+\w^2_k}{2}\right)\tau_S\right]}\right)
\end{align}
where the visibility $V_{jk}$ is given by

\begin{align}
	V_{jk}=\left(\int\ud\w\phi^{1*}_j(\w)\phi^2_j(\w)\right)\left(\int\ud\w\phi^{1*}_k(\w)\phi^2_k(\w)\right).
\end{align}

We can maximize this visibility by maximizing the overlap $f_1$ and $f_2$. We see that this latter provides a lower bound on $V_{jk}$ by writing

\begin{align}
	\int\ud\w\phi^{1*}_j(\w)\phi^2_j(\w)= \frac{\int\ud\w f_1^*(\w,\W_j)f_2(\w,\W_j)}{\sqrt{\rho_1(\W_j,\W_j)\rho_2(\W_j,\W_j)}} \nonumber \\ \geq \int\ud\w\ud\W f_1^*(\w,\W)f_2(\w,\W),
\end{align}
and likewise for $k$.

It is relatively straightforward to maximize the quantity on the left by tuning experimental parameters, namely pump wavelength, phasematching angle, and transverse optical fiber position (due to residual spatial chirp), and observing two-fold coincidences resulting from \text{first order} interference of the sources. Because both sources are pumped with the same pulse, the two-photon term of the state is given by

\begin{align}
	\ket{\psi}\propto \int \ud\w\ud\W \Big (f_1(\w,\W)\adag_1(\w)\bdag_1(\W) + \nonumber \\ f_2(\w,\W)\adag_2(\w)\bdag_2(\W) \Big) \vac.
\end{align}
A straightforward calculation shows that the probability of a two-fold coincidence between ports $\hat{c}$ (or $\hat{d})$ and $\hat{x}$ (or $\hat{y}$) is given by

\begin{align}
	P_{cc} = \frac{1}{4}\int \ud^2\w \Big|f_1(\w,\W) \pm f_2(\w,\W)\Big|^2 \nonumber \\
	=\frac{1}{2}\Big(1 \pm \mathrm{Re}\int \ud^2\w f_1^*(\w,\W)f_2(\w,\W)\Big) \label{eq:supp:ccfringes}
\end{align}

\begin{figure}[t]
	\centering
	\includegraphics[width=.85\linewidth]{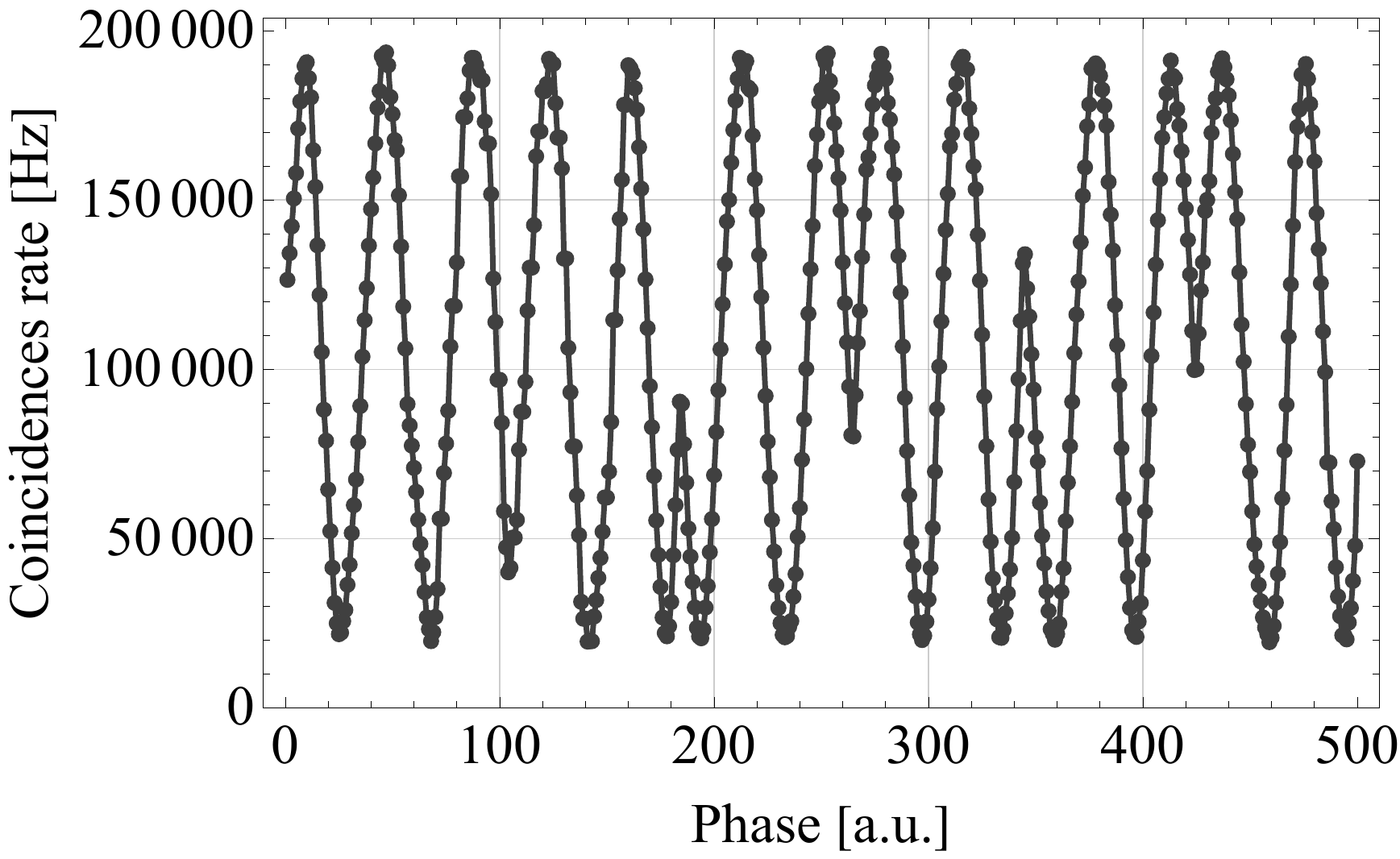}
	\caption{Measured coincidence fringes $P_{\textit{cc}(\tau)}$ with a contrast of $80\%$.}
	\label{fig:supp:coincfringes}
\end{figure}

In the following, we will outline additional measurements to quantify the source indistinguishability. In our case, our dual-pass geometry implies that we need to match the JSI of both sources, which is achieved when both signals and idlers from both sources have maximum overlap. We opted for a bulk crystal source in Type II to enable pumping in both directions while being able to separate our four photons into different paths. We used a BiBO crystal due to its relatively high non linearity.

First, we measured the JSI by directing the two daughter photons from either source into the fiber spools, since their large spectral bandwidth would be cropped with the CFBG. The JSI from each source is depicted in Fig\ref{fig:supp:JSI}. They show that both sources are nearly indistinguishable; a singular value decomposition yields a Schmidt number of $K_1 = 2.9 \pm .1$ and $K_2 = 2.9 \pm .1$. These values are lower than the theoretical expectation ($K\sim 5$) because of the timing jitter of our detectors that result in a broader distribution. This was confirmed by measuring the JSI with the CFBG's, which have a better resolution but are limited in range. The correlation width was found to be lower and therefore the Schmidt number can be expected to be at least $K=4$.

\begin{figure*}[t]
	\centering
	\includegraphics[width=.9\linewidth]{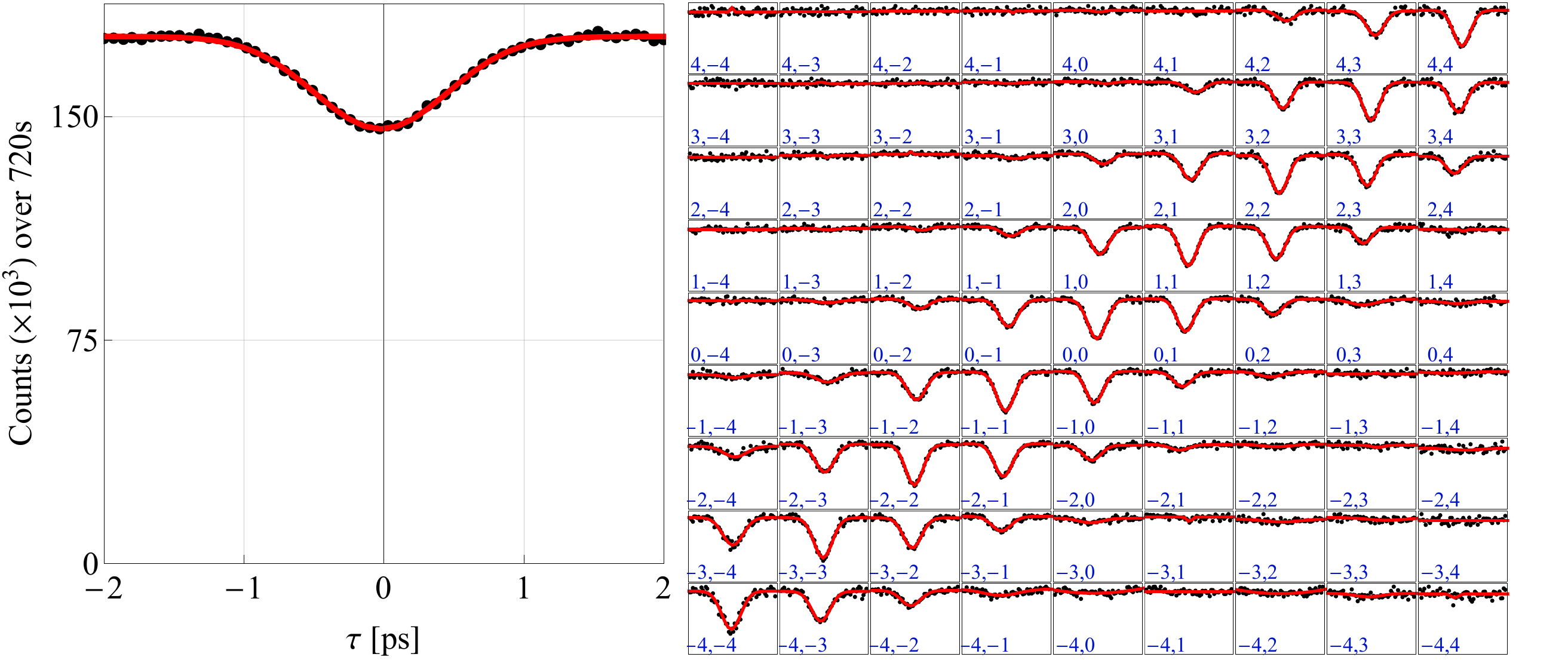}
	\caption{Left: HOM dip between the signal photons heralded by a coincidence between the idler photons. Right: same measurement but with spectral resolution of the heralding photons, labelled $j,k$ for $\W_k,\W_k$, where index $j,k=0$ corresponds to the center frequency $\w_0$.}
	\label{fig:supp:homdip}
\end{figure*}

Note that this method is insensitive to any spectral phase difference, such as dispersion from the pump, since the second pump is slightly more dispersed than the first due to propagation. This has been shown to increase the entanglement and the Schmidt number \cite{Davis2020,Ansari2018}. However, this difference should be negligible, and the method presented that relies on Eq.\eqref{eq:supp:ccfringes} allows for a more accurate estimation of the overlap. Nevertheless, the JSI measurement showed near-perfect correspondence between the intensity of the two sources which is a critical step to ensure indistinguishably between the uncorrelated photon pairs.

To further characterize the indistinguishability of the sources, we measure their heralded $g^{(2)}$ by splitting their signal photon into a beamsplitter. This yields a value of $g_1^{(2)} = 0.16 \pm 0.003$ and $g_2^{(2)} = 0.14 \pm 0.003$. These values are consistent with the relatively high optical power that is utilized to pump the sources in order to maximize the probability of four-fold coincidences. The lower value of $g^{(2)}$ for source 2 is consistent with the fact that it also has a higher heralding efficiency than source 1. The reason is not entirely clear, but it is likely that the previous interaction with the PDC crystal on the first pass results in an additional filtering on the pump as well as a slight reduction in optical power because of absorption.

Finally, in Fig\ref{fig:supp:coincfringes} we measured the coincidences between ports $\hat{c}$ and $\hat{x}$ (see Fig\ref{fig:supp:simple:scheme}) while scanning the relative phase between the two pump fields with a piezoelectric stack, which is related to the probability from Eq.\eqref{eq:supp:ccfringes}. We scanned using a slow voltage ramp resulting in a few micrometers of displacement over a few seconds. The visibility of those fringes is $80\%$, which is a direct measurement of the overlap between the two sources, and therefore a quantification of distinguishability.

Note that we also performed this measurement with spectral resolution, essentially measuring those interferences in narrower spectral bins using our time-of-flight spectrometer. This results in interferences over a much narrower bandwidth, therefore restricting the degrees of freedom of the single photon mode-function. Notably this method has the advantage of being less sensitive to higher order phase mismatch between both sources, such as dispersion. The measured contrast across all spectral bins was found to be $90\%$.

\section{Purity of the heralded states} \label{app:purity}

Since the state $\ket{\psi_{12}}$ from the sources is assumed to be a pure state, the purity of the heralded states $\ket{\Psi_{jk}}$ is ultimately dependent on the amount of spectral filtering in the heralding BSM. To assess this purity, we measure HOM interference between the heralded signal photons when there is no beamsplitter in the idler arms. In this case, upon a coincidence detection of the idler photons at $(\Omega_j,\Omega_k)$, the reduced state of the signal photons is separable, and given by

\begin{align}
	\hat{\rho}_j\otimes\hat{\rho}_k =
	\left(\int \ud^2\omega \rho_j(\omega,\omega')\right)\left(\int \ud^2\tilde{\omega}\rho_k(\tilde{\omega},\tilde{\omega}')\right) \nn \\
	\adag_1(\omega)\adag_2(\tilde{\omega})\ket{\text{vac}}\bra{\text{vac}}\hat{a}_1(\omega')\hat{a}_2(\tilde{\omega}'),
\end{align}
where
\begin{figure*}[t]
	\centering
	\includegraphics[width=\linewidth]{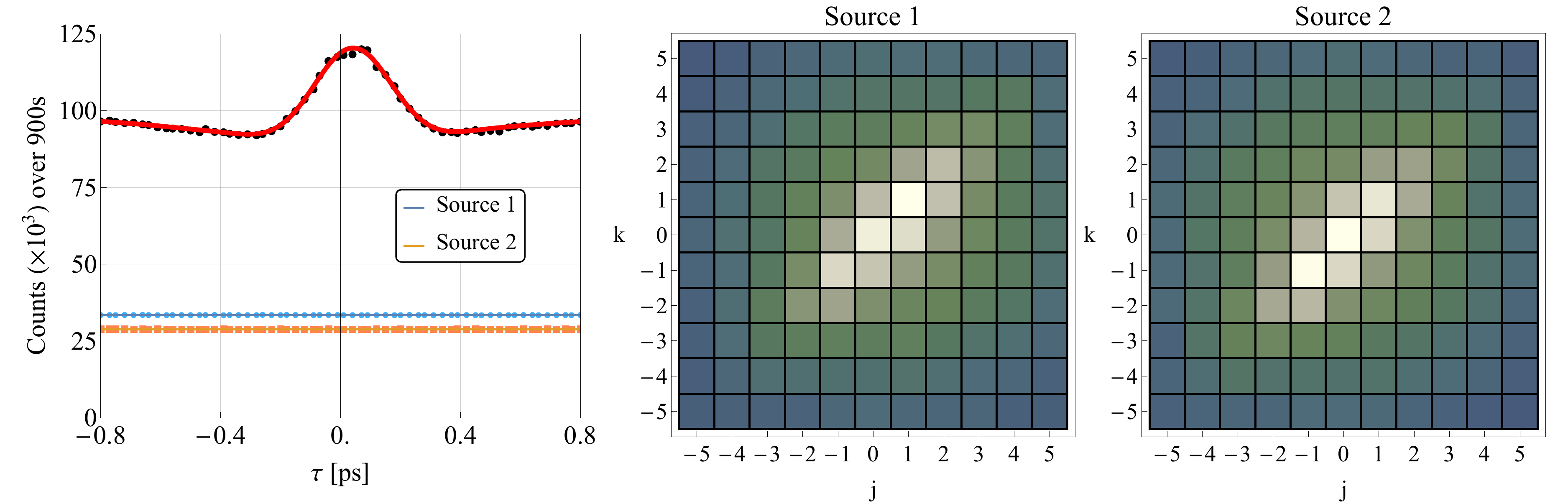}
	\caption{Left: $P(\tau)$ without removing the constant two photon contribution from source 1 (dot) and source 2 (square). $P(\tau)$ resembles Eq.\eqref{eq:Pfull:nodelay} and the fit is obtained by summing the individual fits of $P_{jk}$ as given by \eqref{eq:fullptau:theory}. Right: distribution of these background terms as a function of the heralding frequencies $\W_j,\W_k$}
	\label{fig:supp:background}
\end{figure*}
\begin{align}
	\rho_{j(k)}(\omega,\omega') = \int \ud\Omega |t_{j(k)}(\Omega)|^2 f(\omega,\Omega)f^*(\omega',\Omega).
\end{align}

When the signal photons in this state are incident on a 50:50 beamsplitter, the expected visibility of the HOM interference is given by \cite{Mosley2008}

\begin{align}
	V=\mathrm{Tr}(\hat{\rho}_j\hat{\rho}_k),
\end{align}
and when the idlers are detected in identical frequency bins $(j=k)$, this becomes

\begin{align}
	V=\mathrm{Tr}(\hat{\rho}_{j(k)}^2)=\mathcal{P}(\hat{\rho}_{j(k)}),
\end{align}
where $\mathcal{P}(\cdot)$ denotes the purity of a state. Thus, for $(j=k)$ the visibility of the HOM dip gives a lower bound on the purity of the state $\hat{\rho}_{j(k)}$, and by extension, the state $\hat{\rho}_{jk}$. Our measurements, shown in Fig.~\ref{fig:supp:homdip}, indicate that purity of the heralded states is \textit{at least} 70\%, as evidenced by the HOM visibility along the $j=k$ line. By comparison, a direct calculation of the expected purity using our experimental parameters gives $\sim 78\%$. The purity of our heralded state seems to be dominated by the spectral resolution of our spectrometer. Without spectral resolution, the purity of the heralded state is about 20\% as shown in Fig.~\ref{fig:supp:homdip}.

\section{Background signal} \label{app:background}

As shown by Eq.\eqref{eq:psip}, the full four photon state in the interferometer (see Fig.~\ref{fig:supp:simple:scheme}) contains a contribution from photon pairs emitted by individual sources due to the stochastic nature of parametric down conversion. These terms contribute to $P(\tau_S)$ in the form of interferences that get averaged over the course of a measurement. It is therefore possible to remove that contribution from the signal subsequently to the measurement by blocking a source and recording the rate of four-fold coincidences.

We therefore repeated the measurement of $P_{jk}(\tau_S)$ with either source blocked to obtain the constant background signal for each $j,k$ frequencies, as shown in Fig.~\ref{fig:supp:background}. This shown that the background terms are similar between both sources, therefore the two sources are similar. Summing over all the bins, we can plot on the same scale the contribution of all term in Fig.~\ref{fig:supp:background}. The peak corresponds to interferences from $\ket{\psi_{12}}$ while the flat terms represent $\ket{\psi_{11}}$ and $\ket{\psi_{22}}$. As expected from the theory, both source contribute to $1/4$ of the full signal. Removing those backgrounds at $\W_j,\W_k$ from $P_{jk}$, we obtain the fringes from the main paper with optimal visibility.

\begin{figure}[t]
	\centering
	\includegraphics[width=\linewidth]{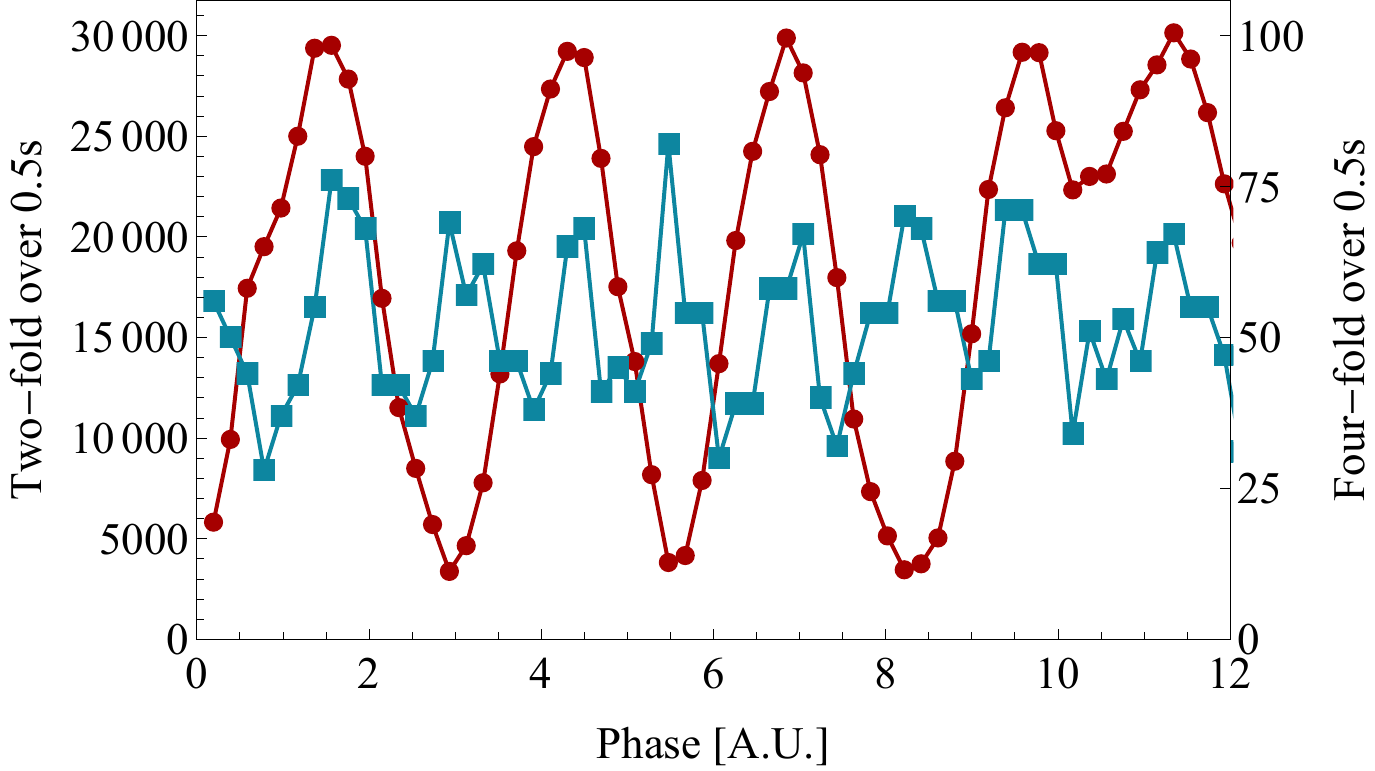}
	\caption{Interference fringes in the two-fold (red) and fourfold (blue) coincidences, obtained in ''real-time", while scanning a PZT as described in the text. The four-fold fringes can be seen to occur at twice the frequency of the two-fold fringes.}
	\label{fig:supp:fastfringes}
\end{figure}

As stated in the main text, we have assumed that the three terms in the full state Eq. \eqref{eq:psip} are mutually incoherent. This is because our measurements are taken over a long timescale of a few hours where the optical phase drifts significantly and any phase-sensitive interference can be neglected. Over a shorter time-scale, we can measure this optical phase in real time by measuring fourfold coincidences with the two beamsplitters present, while scanning the PZT between both sources. A straightforward calculation taking into account the full state \eqref{eq:psip} shows that there is a term that oscillates at the sum frequency $\w+\W \approx \w_p$, where $\w_p$ is the pump frequency, corresponding to about $415$ nm in wavelength. In Fig. \ref{fig:supp:fastfringes}, we plot the measured interference of the two-fold (red) coincidences against the four-fold fringes (blue), where the latter can be seen to modulate at twice the frequency of the two-fold modulation.

\section{Orthogonal modes} \label{app:orthomodes}

From Eq.\eqref{eq:supp:heraldedstate:delay}, we see that the heralded state $\ket{\Psi_{jk}}$ is dependent on the modes $\ket{\phi_j}$ and $\ket{\phi_k}$, which, in the pure state case, results in a heralded joint spectrum \eqref{eq:supp:heraldedJSI:delay} dependent on the outer products $\phi_j(\w_1)\phi_k(\w_2)$. For each heralding bin $j$ and $k$, we label the heralded JSI from \eqref{eq:supp:heraldedstate:delay} as $F_n(\w_1,\w_2)$, where $n$ indexes a pair $(j,k)$. These are normalized as $\int \ud^2\w F_{n}(\w_1,\w_2) = 1 ~ \forall ~ n$ but are not orthogonal, even in the pure state case, \textit{i.e} 
$\int \ud^2\w F_n(\w_1,\w_2)F_m(\w_1,\w_2)  \neq \delta_{nm}$. Orthogonality is usually a corner stone in any quantum protocol, and it is therefore necessary to select the heralded states from our measurement that are orthogonal. To do so, we utilize our measurement of $F_{jk}$, obtained by measuring the spectral coincidences between the signal's photon heralded by a BSM on the idlers. We then obtain a figure similar to Fig.~\ref{fig:supp:sim:JSI} albeit without perfect spectral resolution, putting us in the mixed state configuration, but the strategy to select orthogonal modes within this set is similar to the pure state model.

\begin{figure*}[t]
	\centering
	\includegraphics[width=\linewidth]{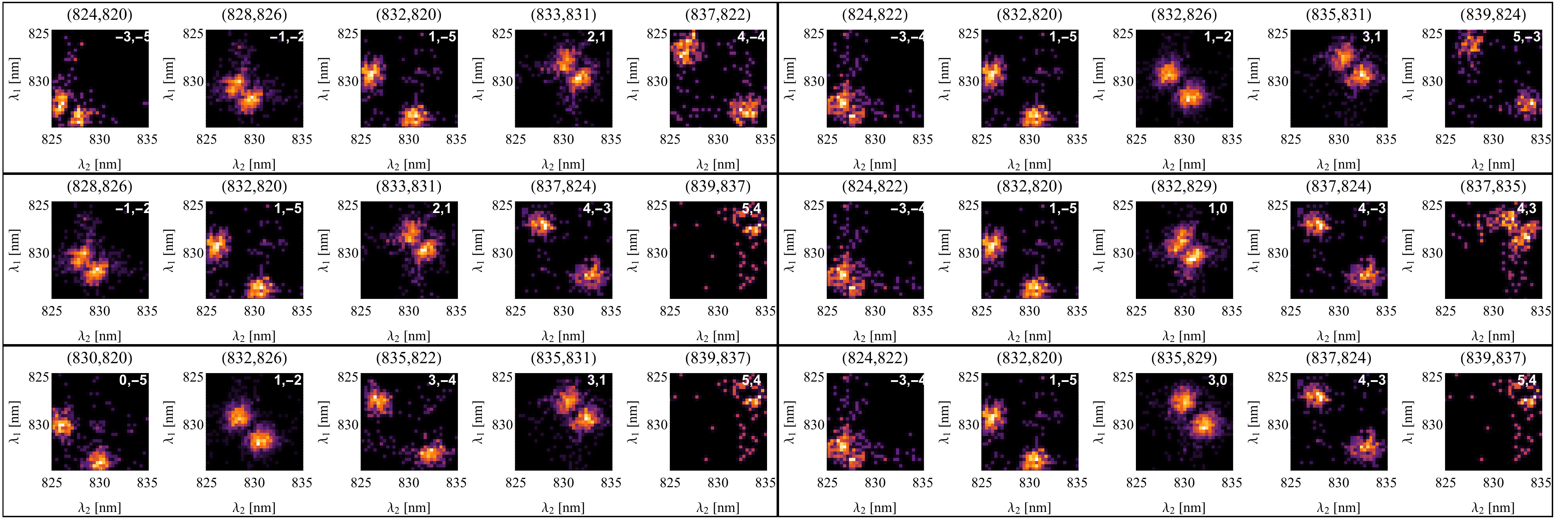}
	\caption{Sets of orthogonal modes $F_{jk}$ that have less than 15 \% of mutual overlap within each part of the grid. The insets label $j$ and $k$.}
	\label{fig:supp:orthomode}
\end{figure*}

First, it is important to notice the symmetry in \eqref{eq:supp:heraldedstate:delay}, where $F_{jk} = F_{kj}$ for $j\neq k$. Since our TOFS are well-calibrated, it is reasonable to symmetrize our measured heralded JSI by averaging the experimentally obtained $F_{jk}$ and $F_{kj}$ (for $j\neq k$) thus defining the $F_n$ functions. Then we compute the mutual overlaps $ \int \ud^2\w F_n(\w_1,\w_2)F_m(\w_1,\w_2) $ and use an algorithm to select a set of modes $\left\{F_{n}\right\}$ which all have an overlap below a certain threshold of $15\%$. We represented a few of these JSI in Fig.~\ref{fig:supp:orthomode}. Since the spectral range of our high resolution TOFS is limited, so is the range over which we can compute overlap, as can be seen from the modes that are labelled with a large $j,k$. Nevertheless, there is a sufficient amount of spectral coincidence in those cases to infer orthogonality with the other JSI.

Note that while this overlap is computed between the joint spectral intensities and not between the states, it can be shown that if the overlap in intensity is zero, then the states are necessarily orthogonal, hence the strategy is valid to select which $\ket{\Psi_{jk}}$ are mutually orthogonal. Therefore, it is fair to say that the JSI $F_{jk}$ from Fig.~\ref{fig:supp:orthomode} correspond to the heralded states $\ket{\Psi_{jk}}$ that are all mutually orthogonal.

\section{Miscellaneous functions and relations} \label{app:math}


Our theoretical derivation relies on the definition of the $\phi_{j(k)}(\w)$ functions which renders computation easier thanks to the Gaussian approximation. These functions can be also written in a density matrix formalism.\\
Using the definitions from Sec.\ref{sec:theory}, the $\phi_{j(k)}$ functions are defined from the JSA by:
\begin{align}
    f(\w,\W_{j(k)} = \sqrt{N_j} \phi_{j(k)}(\w),
\end{align}
where $N_j$ is a function that depends on the heralding frequency $\W_j$, the JSA bandwidth over the idler axis $\sigma_\text{I}$ and the amount of entanglement $\alpha$.
The reduced density matrix of the idler, given by Eq.\eqref{eq:rhoId}, can then be written in the following manner:
\begin{align}
    \rho_\mathrm{I}(\W_j,\W_k) &= \int \ud \w \ f(\w,\W_j) f^\ast(\w,\W_k) \\
    &= \sqrt{N_j N_k} \int \ud \w \ \phi_j(\w) \phi_k^\ast(\w)\\
    &= \sqrt{N_j N_k} \braket{\phi_j | \phi_k},
\end{align}
where we see that the idler density matrix can be linked to the overlap integral between the heralded signal states. The $N_{j(k)}$ functions are then found to be equal to the diagonal elements of the idlers density matrix:
\begin{align}
    N_{j(k)} = \rho_\mathrm{I}\left( \W_{j(k)},\W_{j(k)} \right),
\end{align}
since the $\phi_{j(k)}$ are $\ell^2$ normalized.

The signal density matrices then follow a similar derivation, with
\begin{align}
    \rho_\mathrm{S}(\w,\w') &= \int \ud \W_{j(k)} \ f(\w,\W_{j(k)})f^\ast(\w',\W_{j(k)}) \\
    &= \int \ud \W_{j(k)} \ N_{j(k)} \phi_{j(k)}(\w) \phi_{j(k)}^\ast(\w),
\end{align}
and the diagonal elements are given by
\begin{align}
    \rho_\mathrm{S}(\w,\w) = \int \ud\W_{j(k)} \ N_{j(k)} \modsqr{\phi_{j(k)}(\w)}.
\end{align}
It is easy to show that in the case of similar sources, we also have the identity
\begin{align}
    \rho_\mathrm{S}(\w,\w')=\rho_\mathrm{S}^\ast(\w',\w).
\end{align}
Finally, the overlap between the modes $\phi_j$ and $\phi_k$ can also be written in term of the density matrices, as:
\begin{align}
    \braket{\phi_j | \phi_k} = \frac{\rho_\mathrm{I}(\W_j,\W_k)}{\sqrt{N_j N_k}}
\end{align}

\end{appendix}

\bibliography{biblio}

\end{document}